\documentclass[prd,aps,twocolumn,a4paper,floatfix,showpacs,nofootinbib]{revtex4-1}

\usepackage[utf8]{inputenc}  
\usepackage[T1]{fontenc}
\usepackage{graphicx,psfrag}
\usepackage{mathrsfs}
\usepackage{amsmath,amsfonts,amssymb,gensymb}
\usepackage{multirow,enumerate}
\usepackage{comment,hyperref}
\usepackage{color}
\usepackage{acronym}
\usepackage{xspace}
\usepackage[normalem]{ulem}
\usepackage{mathtools}
\usepackage{subfigure}
\usepackage{makecell}
\usepackage{appendix}

\newacro{BH}{black hole}
\newacro{NS}{neutron star}
\newacro{PN}{Post-Newtonian}
\newacro{BBH}{binary black hole}
\newacro{BNS}{binary neutron star}
\newacro{EOB}{effective-one-body}
\newacro{NR}{numerical relativity}
\newacro{GW}{gravitational wave}
\newacro{EOS}{equation-of-state}

\newcommand{\be}{\begin{equation}}
\newcommand{\ee}{\end{equation}}
\newcommand{\bea}{\begin{eqnarray}}
\newcommand{\eea}{\end{eqnarray}}
\newcommand{\bel}{\begin{align}}
\newcommand{\eel}{\end{align}}

\def\GMc2{{\rm G M_{\odot} c^{-2}}}

\def\SEOBNRv4T{\texttt{SEOBNRv4T}\xspace}

\usepackage{color}
\definecolor{cyan}{rgb}{0,0.9,0.9}
\definecolor{orange}{rgb}{0.9,0.5,0}
\definecolor{magenta}{rgb}{1,0,1}
\definecolor{purple}{rgb}{0.8,0.4,0.8}
\definecolor{gray}{rgb}{0.5,0.5,0.5}
\definecolor{mygreen}{rgb}{0.1,0.8,0.1}
\definecolor{darkblue}{rgb}{0.0,0.0,0.6}

\begin{document}

\title{Biases in parameter estimation from overlapping 
gravitational-wave signals in the third generation detector era} 

\author{Anuradha Samajdar$^{1,2,3}$}
\author{Justin Janquart$^{1,2}$}
\author{Chris Van Den Broeck$^{1,2}$}
\author{Tim \surname{Dietrich}$^{4,5}$}

\affiliation{${}^1$Nikhef, Science Park 105, 1098 XG Amsterdam, The Netherlands}
\affiliation{${}^2$Institute for Gravitational and Subatomic Physics (GRASP), 
Utrecht University, Princetonplein 1, 3584 CC Utrecht, The Netherlands}
\affiliation{${}^3$Department of Physics, University of Milano -- Bicocca, 
Piazza della Scienza 3, 20126 Milano, Italy}
\affiliation{${}^4$Institut f\"{u}r Physik und Astronomie, Universit\"{a}t Potsdam, Haus 28, 
Karl-Liebknecht-Str. 24/25, 14476, Potsdam, Germany}
\affiliation{${}^5$Max Planck Institute for Gravitational Physics (Albert Einstein Institute), Am M\"uhlenberg 1, Potsdam 14476, Germany}

\date{\today}

\begin{abstract}
In the past few years, the detection of gravitational waves from compact binary 
coalescences with the Advanced LIGO and Advanced Virgo detectors has become routine. 
Future observatories will detect even larger numbers of gravitational-wave signals, 
which will also spend a longer time in the detectors' sensitive band. 
This will eventually lead to overlapping signals, especially in the 
case of Einstein Telescope (ET) and Cosmic Explorer (CE). Using realistic distributions
for the merger rate as a function of redshift as well as for component masses in binary neutron star and 
binary black hole coalescences, we map out how often signal overlaps of various types will occur in an ET-CE
network over the course of a year. We find that a binary neutron star signal will typically have 
tens of overlapping binary black hole and binary neutron star signals. Moreover, it will happen up to 
tens of thousands of times per year that two signals will have their end times within seconds of each 
other.  
In order to understand to what extent this would lead to measurement biases 
with current parameter estimation methodology, we perform injection studies with 
overlapping signals from binary black hole and/or binary neutron star coalescences. 
Varying the signal-to-noise ratios, the durations of overlap, and the kinds of overlapping signals, we 
find that in most scenarios the intrinsic parameters can be recovered
with negligible bias. However, biases do occur for a short binary black hole or a quieter binary neutron 
star signal overlapping with a long and louder binary neutron star event when the merger times are 
sufficiently close. Hence our studies show where improvements are required to ensure reliable estimation of 
source parameters for all detected compact binary signals as we go from second-generation
to third-generation detectors.
\end{abstract}

\maketitle

\section{Introduction}
\label{sec:intro}
The direct observation of gravitational waves (GWs)~\cite{TheLIGOScientific:2016htt} has 
had a tremendous impact in fundamental physics 
\cite{TheLIGOScientific:2016src,LIGOScientific:2018jsj,Abbott:2018lct,LIGOScientific:2019fpa}, 
astrophysics 
\cite{Abbott:2016nhf,LIGOScientific:2018mvr,Abbott:2020niy,TheLIGOScientific:2017qsa,Abbott:2018exr,Abbott:2020uma,LIGOScientific:2020stg,Abbott:2020khf,Abbott:2020tfl,Abbott:2020mjq}, 
and cosmology \cite{Abbott:2017xzu,Soares-Santos:2019irc,Palmese:2020aof}, and starting from the 
observation of the binary neutron star (BNS) signal GW170817 \cite{TheLIGOScientific:2017qsa} has opened a 
new era in multi-messenger astronomy with GWs \cite{Soares-Santos:2017lru,Cowperthwaite:2017dyu,Monitor:2017mdv,ANTARES:2017bia}. 
The third observing run (O3) of Advanced LIGO~\cite{TheLIGOScientific:2014jea} and 
Advanced Virgo~\cite{TheVirgo:2014hva} ended in March 2020, and together these 
interferometers have found more than 50 GW candidates~\cite{gracedb}, with 39 candidates observed 
during the first half of O3~\cite{Abbott:2020niy}. 
The detector sensitivities will be improved further, and the frequency with which signals 
are observed is expected to keep increasing in coming years. 
In particular, in the transition to the envisaged third generation (3G) era, with  
Einstein Telescope (ET)~\cite{Punturo_2010,Hild_2011} and Cosmic Explorer 
(CE)~\cite{Reitze:2019iox, Abbott_2017, Regimbau_2017}, the detection rate will go up steeply,  
and signals will also spend much longer times in the detectors' sensitive band~\cite{Sathyaprakash:2012jk}.  
As first pointed out in \cite{Regimbau:2009rk} and studied further in this paper, the probability of 
overlapping signals will then become signficant. 

In view of this, it will be important to assess 
to what extent the science goals of 3G detectors 
(see e.g.~\cite{Sathyaprakash:2009xt,VanDenBroeck:2010vx,Zhao:2010sz,Punturo:2010zza,Sathyaprakash:2011bh,Sathyaprakash:2012jk,Broeck:2013rka,Maggiore:2019uih,Sathyaprakash:2019yqt,Vitale_2019,ng2020probing})
may be affected by signals overlapping with each other. Apart from science with signals from 
compact binary coalescences (CBCs), this includes searches for primordial backgrounds, since 
the subtraction of ``foreground'' CBC sources 
\cite{Cutler_2006,Harms_2008,Regimbau_2017,Sachdev_2020,Sharma_2020,Biscoveanu_2020} 
will rely on our ability to characterize them individually. 
As shown in \cite{Regimbau:2012ir,Meacher:2015rex}, even using current data analysis techniques, the 
\emph{detection rates} of individual CBC sources would likely not be significantly impacted by the
occurrence of overlapping signals. 
However, a study of the effect on \emph{parameter estimation} had not yet been
performed. 

Earlier works~\cite{Vitale_2016, Vitale_2017, Vitale_2018} have studied parameter estimation for 
single sources in the 3G era. Here we take the first step in assessing possible biases in the recovery of 
parameters characterizing a GW signal when signals from different sources are simultaneously 
present in the detectors' sensitive band. 
Before doing this, we map out how often signal overlaps of 
various types will occur in a network of two CEs and one ET over the course of a year, assuming realistic distributions
for merger rate as a function of redshift and for component masses in binary neutron star and 
binary black hole (BBH) coalescences. We find that a typical BNS signal will be overlapped by tens of 
BBH signals. Moreover, BBH or BNS signals whose mergers occur within seconds from each other will be quite common. 
Since these are the cases for which we can expect the largest parameter estimation biases to occur, we focus on them 
in setting up simulations whereby signals are added to synthetic data from the ET-CE network, 
and analyzed using current state-of-the-art parameter estimation techniques. 
We explore various scenarios of signals from different kinds of 
sources overlapping: two BBH signals, two BNS signals, and a BBH signal with a BNS. 
For our simulations we choose signal parameters consistent with what has been observed as being 
representative of each kind of source: parameter values similar to the ones of 
GW170817~\cite{Abbott:2018wiz} to represent a BNS, similar to the ones of GW150914~\cite{Abbott:2016blz} 
to represent a high-mass BBH, and similar to the ones of GW151226~\cite{Abbott:2016nmj} 
for a lower-mass BBH. We find that in most cases, the intrinsic parameters can 
be recovered with negligible bias. However, if the merger times of the two signals are 
sufficiently close, considerable biases can occur when a short BBH
signal or a quieter BNS signal overlaps with a louder BNS signal.  
Though our study should be considered exploratory, it already points to where improvements 
over current parameter estimation pipelines will be needed the most.

This paper is structured as follows. In Sec.~\ref{sec:rates} we obtain detection rate estimates for 
signals in the 3G era, from which we calculate overlap rates. In Sec.~\ref{sec:method}  
we lay out the settings and methods we use for parameter estimation. 
Parameter estimation results for various scenarios are shown in Sec.~\ref{sec:results}.  
A summary and conclusions are presented in Sec.~\ref{sec:conclusion}, where we also give recommendations
for future improvements of parameter estimation techniques.

\section{Overlap rate estimates}
\label{sec:rates}
\subsection{Methodology}

Before looking at the impact of overlapping signals on parameter estimation for the individual ones, 
we want to address the question of 
how frequently such overlaps will occur, depending on the type. 
Previous characterizations of the overlap probabilities for 3G detectors were based on 
the \emph{duty cycle}, which is defined as the ratio of 
the typical duration of a particular type of event (BNS or BBH) to 
the average time interval between two successive 
events of that type, assuming some fixed canonical values of the component masses 
for each type \cite{Regimbau:2009rk}. However, here we also want to allow for overlaps of mixed types, 
and for a range of component masses (and hence signal durations) within a given type, so as to arrive
at a detailed assessment of overlap rates. Therefore, what we will do is to assume particular merger 
rates as function of redshift for BBH and BNS, as well as component mass distributions, and on the basis of 
these create simulated ``catalogs'' of signals in the detectors. This will allow us to make 
quantitative statements regarding BNS signals overlapping with other BNS signals and with BBHs, and 
the same for overlaps of BBH with BBH events, in a much more detailed and realistic 
fashion.\footnote{Since neutron star-black hole (NS-BH) rates are less certain (see 
e.g.~\cite{Abbott:2020khf,Hoang:2020gsi}), we will not consider
them here, but we expect general conclusions regarding parameter estimation to largely carry over when signal
durations are similar.}

\vspace{5mm}

\begin{figure}
    \centering
    \includegraphics[scale = 0.6]{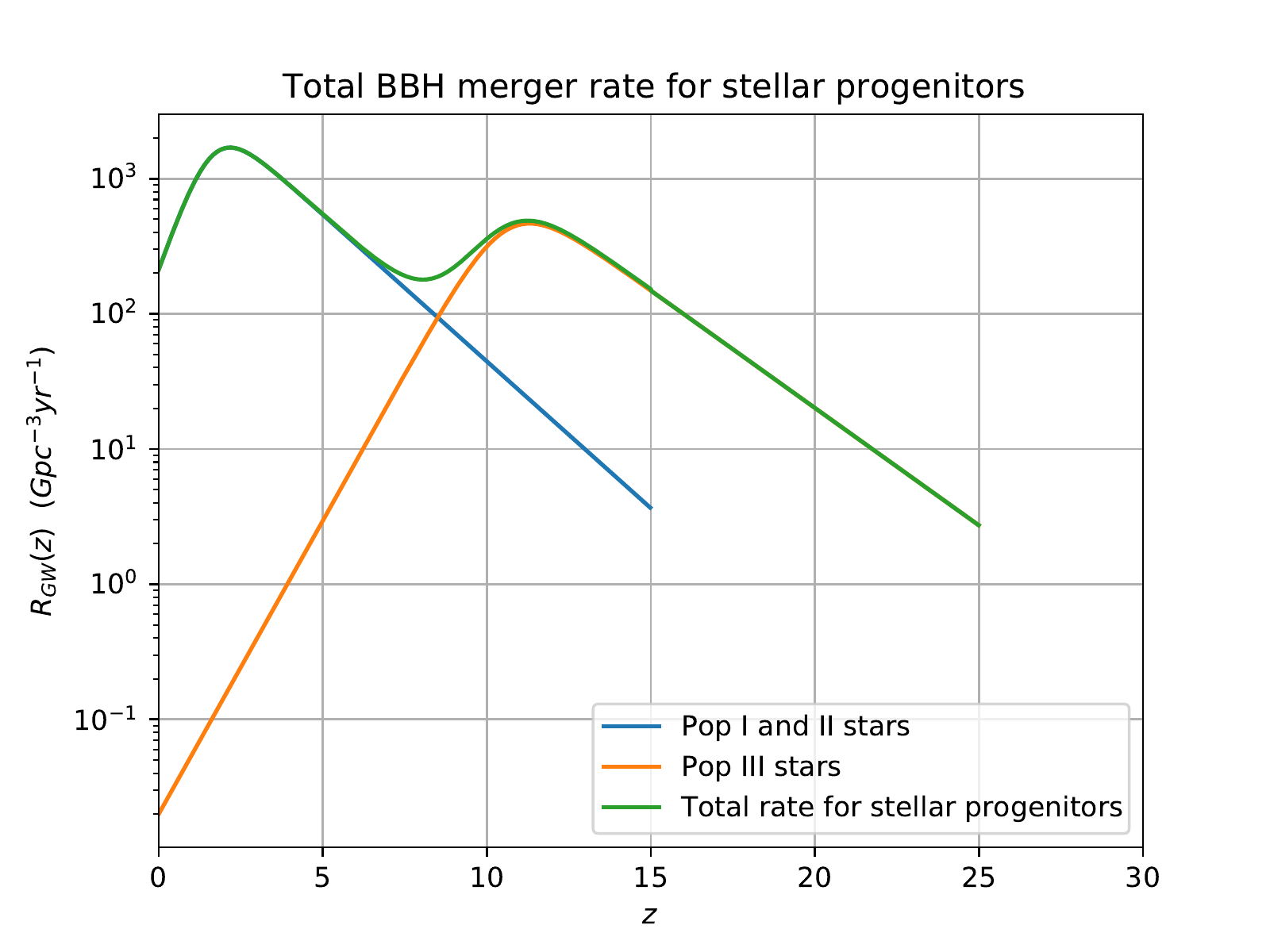}
    \caption{The BBH merger rate density according to Oguri's fit \cite{Oguri:2018muv} 
    for population I, II, and III stars, as well as the total rate, when all the star 
    populations are accounted for.}
    \label{fig:OguriFit}
\end{figure}

We start by estimating the number of individual BBH and BNS 
coalescences that happen in a given volume, up to a maximum redshift which is chosen to be 
$z_{\rm max} = 30$ for BBH events and $z_{\rm max} = 6$ for BNS events  
\cite{Regimbau:2009rk,Regimbau:2012ir,Sathyaprakash:2012jk,Safarzadeh:2019pis}. 
For this we need the intrinsic merger rate density for the events 
as a function of redshift. We will assume that the compact binaries originate from stellar populations,  
and adopt the merger rate estimates of Belcynski et al.~\cite{Belczynski:2016ieo} 
with Oguri's analytical fit \cite{Oguri:2018muv}\footnote{Strictly speaking this merger 
rate distribution refers to BBH mergers. However, when computing the 
merger rate density (see e.g.~\cite{Regimbau:2012ir, Belczynski:2016ieo, Safarzadeh:2019pis}), 
one assumes a time delay distribution (e.g.~$P(t_{\rm d}) \propto 1/t_{\rm d}$), with a minimum 
time delay that is higher for BBH than for BNS. Using the distribution of \cite{Belczynski:2016ieo}
for both BNS and BBH (with some overall rescaling) then implies that we will underestimate the BNS merger rate density 
\cite{Safarzadeh:2019pis} and hence the frequency of overlaps involving BNS signals.}, whose
general expression is 
\begin{equation}
R_{\rm GW}(z) = \frac{a_{1}e^{a_{2}z}}{e^{a_{3}z}+a_{4}}.
\end{equation}
Here $R_{\rm GW}$ is expressed in ${\rm Gpc^{-3}\, yr^{-1}}$, and the coefficients $a_i$, 
$i = 1, \ldots, 4$ depend on the star populations that are considered; see Fig.~\ref{fig:OguriFit}. For 
our purposes, we consider the combination of population I and II stars for BNS, and populations 
I, II, and III for 
BBH, as the contribution of the latter type of stars is important only at redshifts of $\gtrsim 4$.  
However, these relations are rescaled to match the local merger rate estimates obtained observationally by
LIGO and Virgo so far; see \cite{Abbott:2020gyp}. 
In this work, we focus on the lowest, the median, and the highest local rate for each type of event. 
For BNS, the lowest, median, and highest local rates are, respectively, $80\,{\rm Gpc^{-3}\,yr^{-1}}$, 
$320\,{\rm Gpc^{-3} \,yr^{-1}}$, and $810\,{\rm Gpc^{-3}\, yr^{-1}}$, which are obtained by 
changing the value of $a_1$ to 2480, 9920, and 25110, respectively. On the other hand, for the BBH events, 
we apply a multiplicative constant to the sum of the population I and II and the population III rates,  
equal to 0.0709, 0.112, and 0.178 for the lowest, median, and highest local rates, which are  
$15.1\,{\rm Gpc^{-3} \,yr^{-1}}$, $23.8\,{\rm Gpc^{-3}\, yr^{-1}}$, and $37.9\,{\rm Gpc^{-3} \,yr^{-1}}$, 
respectively.

An intrinsic merger rate density $R_{\rm GW}(z)$ is then converted to an \emph{observed} 
merger rate density as function of redshift 
by multiplying by the differential comoving volume \cite{Regimbau:2009rk}: 
\begin{equation}
R^{\rm obs}_{\rm GW}(z) = R_{\rm GW}(z)\,\frac{dV_c}{dz}(z).
\end{equation}
To obtain $dV_c/dz$, we assume the \emph{Planck13} cosmological model \cite{Ade:2013zuv} of 
\emph{Astropy} \cite{Astropy2013, Astropy2018}. 

As a next step, we simulate the population of systems by constructing a ``catalog'', 
and determine which events are actually detected. 
For the BBH population, we assume that the masses follow the ``\emph{power law + peak}'' 
distribution presented in Ref.~\cite{Abbott:2020gyp} for the primary component mass, and the 
corresponding power law distribution for the mass ratio, through which we sample the secondary 
mass \cite{Abbott:2020gyp}. For BNS events we distribute component masses uniformly, where 
for the primary mass $m_1 \in [ 1, 2.5 ]\,M_\odot$, and for the secondary mass 
$m_2 \in [ 1\,M_\odot, m_1 ]$. Events are distributed over comoving distance $D$ according to 
$R_{\rm GW}(z)$, converting between $D$ and $z$ using the above mentioned cosmology and cutting off
at the maximum redshifts $z_{\rm max}$ stated above. Sky positions
and unit normals to the orbital plane are taken to be uniform on the sphere.

In this work we assume a network of two CEs located at the LIGO Hanford and Livingston sites, 
and one ET located at the Virgo site. For each event we calculate the optimal signal-to-noise ratios
(SNRs) in the three observatories, which are added in quadrature to obtain a network SNR. 
In computing SNRs we only consider the inspiral part of binary coalescence, so that 
in the stationary phase approximation \cite{Sathyaprakash:1991mt} and for a single interferometer 
\cite{Kastha:2018bar}
\begin{eqnarray}
{\rm SNR} &=& \frac{1}{2}\sqrt{\frac{5}{6}}\frac{1}{\pi^{2/3}}\frac{c}{D(1+z)^{1/6}}
\bigg(\frac{G\mathcal{M}}{c^{3}} \bigg)^{5/6} \nonumber\\
&& \times \,\,g(\theta, \phi, \psi, \iota)\,\sqrt{I(M)}. 
\label{eq:rho}
\end{eqnarray}
Here $\mathcal{M} = (m_{1}\,m_{2})^{3/5}/(m_{1}+m_{2})^{1/5}$ is the chirp mass 
in the source frame. The geometric factor is given by 
\begin{equation}\label{eq:g_ang}
\begin{split}
g(\theta, \phi, \psi, \iota) = \bigg(F_{+}^{2} & (\theta, \phi, \psi)(1+\cos(\iota)^{2})^{2} 
\\ & + 4F_{\times}^{2}(\theta, \phi, \psi)\cos(\iota)^{2}\bigg)^{1/2},
\end{split}
\end{equation}
where $F_{+,\times}$ are the beam pattern functions in terms of sky position $(\theta, \phi)$
and polarization angle $\psi$, while $\iota$ is the inclination angle. We take 
Einstein Telescope to consist of three detectors with $60^\circ$
opening angle, arranged in a triangle with sides of 10 km \cite{Freise:2008dk}, and add the
corresponding SNRs in quadrature; for Cosmic Explorer we assume a single L-shaped detector 
of 40 km arm length \cite{Reitze:2019iox, Abbott_2017}. 
Finally, 
\begin{equation}\label{eq:FreqInt}
I(M) = \int_{f_{\rm low}}^{f_{\rm high}}\frac{f^{-7/3}}{S_{h}(f)}df.
\end{equation}
Here $f_{\rm low}$ is a low-frequency cut-off that depends on the observatory; we 
set $f_{\rm low} = 5$ Hz for both ET and CE, though lower values may be achieved in the
case of ET \cite{Hild:2010id,Hild:2009ns}. For $f_{\rm high}$ we use the frequency of 
the innermost stable circular orbit: 
\begin{equation}\label{eq:fmax}
    f_{\rm high}(m_{1}, m_{2}, z) = \frac{1}{1+z}\frac{1}{6\pi\sqrt{6}}\frac{c^{3}}{GM},
\end{equation}
where $M = m_1 + m_2$ is the total mass. We take the noise power spectral density (PSD) $S_h(f)$ to
be ET-D in the case of Einstein Telescope \cite{Punturo_2010,Hild_2011}; for the projected
PSD of Cosmic Explorer, see \cite{Reitze:2019iox,Abbott_2017}. 

The network SNR, denoted $\mbox{SNR}_{\rm net}$, is defined as 
\begin{equation}\label{eq:NetSNR}
    \mbox{SNR}_{\rm net}^2 = \sum_{i=1}^3 \mbox{SNR}_{i}^{2},
\end{equation}
where the sum is over the two CE and the one (triangular) ET observatories. 
We consider an event as detectable 
if the network SNR is above 13.85 ($= \sqrt{3} \times 8$), without 
imposing SNR thresholds in individual observatories. For the BNS and BBH mass ranges considered here, 
this means that detection rates will mainly be driven by the CEs, but we note that ET will 
have an advantage at higher masses \cite{sathyaprakash2019cosmology}. 

Finally, 
signals will be present in a detector for a duration given by 
\begin{eqnarray}\label{eq:duration}
&&\tau
= 2.18\,\bigg(\frac{1.21\,M_\odot}{\mathcal{M}} \bigg)^{5/3}
\bigg[ \bigg(\frac{100\,\mbox{Hz}}{f_{\rm low}}\bigg)^{8/3}
-\bigg(\frac{100\,\mbox{Hz}}{f_{\rm high}}\bigg)^{8/3}\bigg]\,\mbox{s}.\nonumber\\
&&
\end{eqnarray}

Simulated catalogs of events happening over the course of a year are constructed as follows. The year is 
split into a grid in which each cell corresponds to one second, and merger times are drawn from a uniform 
distribution over these cells. For a given type of event (BNS or BBH), one associates to each merger time 
a mass pair, redshift, sky position, and orientation of the orbital plane drawn from the corresponding distributions, 
as well as a signal duration computed from Eq.~(\ref{eq:duration}). Doing this for the three choices of local merger 
rate, and in each case putting together the BNSs and BBHs, catalogs of events are obtained. Finally, 
within each catalog, it is assessed which events will be detectable with the ET-CE network according to the 
criteria spelled out above, leading to an overview of what we may expect to be contained in one year's
worth of data. In particular, we can check how often and in what way events 
tend to overlap, depending on their types.

\begin{table*}[t]
\begin{footnotesize}
\renewcommand{\arraystretch}{1.2}
		\begin{tabular}{|c|c|c|c|c|c|c|c|}
			\hline
			&  \# of detections & $\mbox{SNR}_{\rm net}$ & \# with $\mbox{SNR}_{\rm net} > 250$ & \# with $\mbox{SNR}_{\rm net} > 100$ & \# with $\mbox{SNR}_{\rm net} > 50$ & \# with $\mbox{SNR}_{\rm net} > 20$\\
			\hline
			\textbf{BBH} & & & & & & \\
			\hline
			Low rate & 53756 & $81.1_{-57.3}^{+94.2}$ & 3069 (5\%) & 20605 (35\%) & 40063 (68\%) & 52239 (89\%) \\
			Median rate &  85725 & $81.3_{-57.5}^{+93.9}$ & 4972 (5\%) & 33148 (39\%) & 63958 (75\%) & 83333 (97\%) \\
			High rate & 137225 & $81.5_{-57.4}^{+94.2}$ & 7860 (6\%) & 53419 (39\%) & 102766 (75\%) & 133460 (97\%) \\
			\hline
			\textbf{BNS} & & & & & & \\
			\hline
			Low rate & 98898 & $19.2_{-4.9}^{+22.1}$ & 17 (0.017\%) & 298 (0.30\%) & 2712 (2.7\%) & 44350 (48\%) \\
			Median rate & 396793 & $19.1_{-4.8}^{+22.0}$ & 73 (0.018\%) & 1257 (0.32\%) & 10659 (2.7\%) & 177296 (45\%) \\
			High rate & 1004525 & $19.1_{-4.8}^{+22.1}$ & 196 (0.020\%) & 3255 (0.32\%) & 27135 (2.7\%) & 448610 (45\%) \\
			\hline
		\end{tabular}
		\caption{The number of events detected by a network of two CEs and one ET  
		in one year of simulated data, the median network SNRs and their 90\% spreads, and the detection 
		numbers and 
		percentages (in brackets) for different choices of minimum network SNR.}
		\label{tab:DetectionsET}
\end{footnotesize}
\end{table*}

\subsection{Overlap estimates}

The three different local merger rates give the following typical numbers of events happening over one year,  
prior to imposing detectability thresholds:  
$\sim 59000$, $93000$, $148000$ BBH events, and $286000$, $1145000$, $2900000$  
BNS events for the low, median, and high local rate, respectively. The network of two CEs and one ET
will detect 93\% of BBHs and 35\% of BNSs. The number of detected signals is shown in Table \ref{tab:DetectionsET}
for the three local rates, along with median and 90\% spreads on SNRs, and a breakdown of detections
according to their loudness. 

Within our simulated catalogs of events, we can look at the numbers 
of detected signals that overlap depending on the types. 
We focus on two quantities: (i) the number of seconds in a year where 
at least two detected signals have their merger, and 
(ii) the typical number of mergers that happen during the time a given signal is in a detector's
sensitivity band. 

The numbers of seconds in a year that have at least two mergers taking place is given in 
Table~\ref{tab:MergSame}; clearly this will happen  
frequently over the course of a year. Indeed, we find that even more than two mergers can occur 
within the same second. The proportion of detected signals merging together with at least one other 
goes up with increasing local merger rate, potentially reaching thousands per year. 

\begin{table}[t]
\begin{footnotesize}
\renewcommand{\arraystretch}{1.2}
	\begin{tabular}{|c|c|c|c|}
		\hline
		Rate & BBH mergers > 1 & BNS mergers > 1   &  Any mergers > 1 \\
		\hline
		Low rate & 48  & 310 & 750  \\
		Median rate & 127  & 2412  &  7347  \\
		High rate & 303  & 15581  & 20149  \\
		\hline
	\end{tabular}
	\caption{ The number of seconds in a year with at least two mergers occurring, depending on their 
	types.}
	\label{tab:MergSame}
\end{footnotesize}
\end{table}

\begin{figure}
    \centering
   \includegraphics[keepaspectratio, width=0.5\textwidth]
   {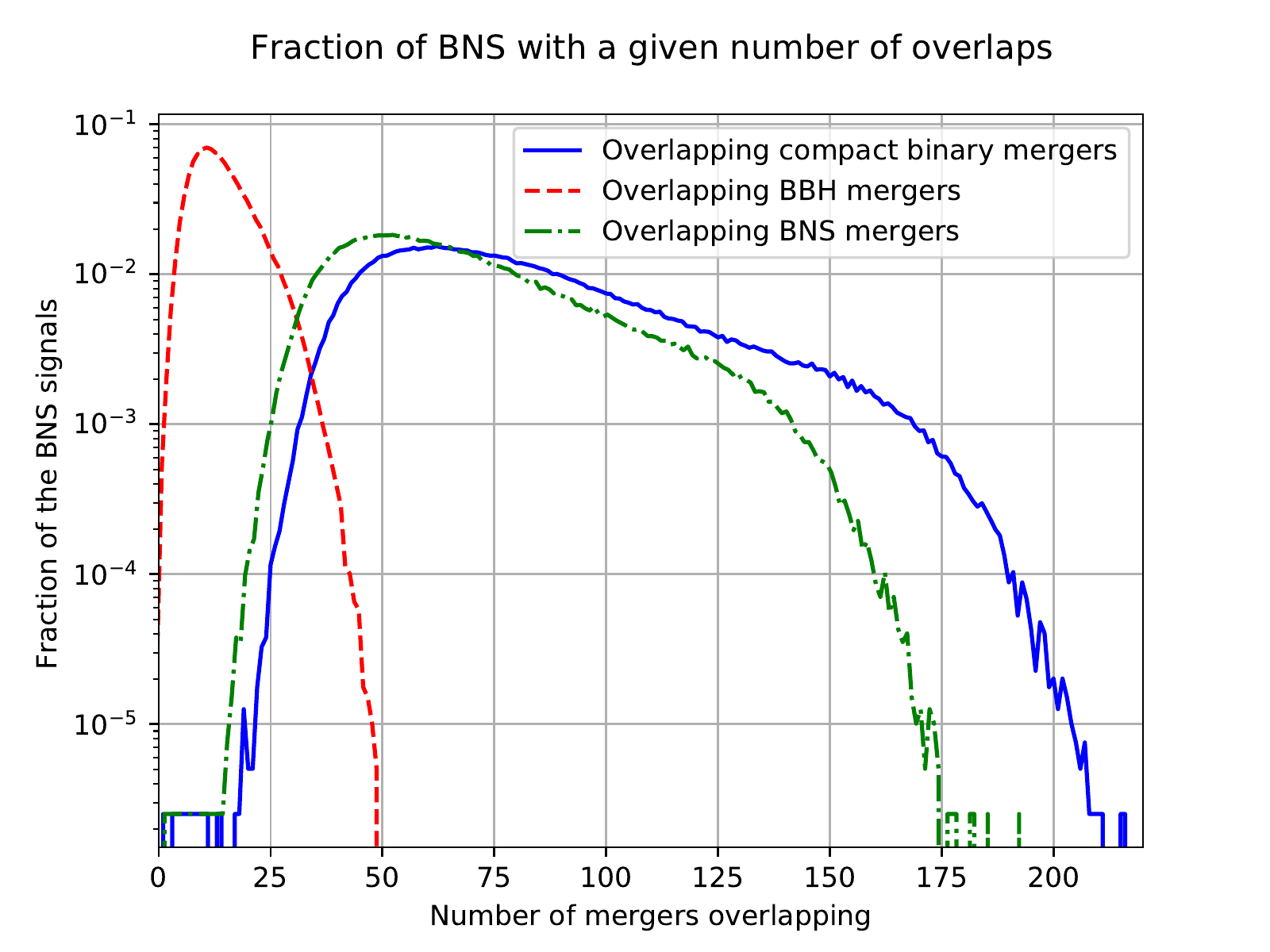}
    \caption[width = 0.8\textwidth]{Fraction of detected BNS mergers with a given number 
    of compact binary mergers (blue), BBH mergers (red), and BNS mergers (green) taking place 
    while the BNS signal is in band.}  
    \label{fig:HistBNS}
\end{figure}

In addition to the scenario where different compact binary mergers happen in the same second, 
we investigate the typical number of mergers that will happen over the entire duration of a BNS event while
it is in band, depending on their type; see Table \ref{tab:BNSover} and Fig.~\ref{fig:HistBNS}. 
Because BNS events are in the detector band for a long time (several hours for $f_{\rm low} = 5$ Hz), 
quite a number of such overlaps will indeed occur. If one does the same for BBHs, one finds that 
either zero or one BBH or BNS merger (at 90\% confidence) will happen in its duration; this is 
due to BBH events being shorter (the median duration being $\sim 45$ seconds).

Before moving on to parameter estimation issues, let us briefly look at other future GW 
observatories that are being planned or considered. Constructing simulated catalogs of 
detectable sources in the same way as above, and focusing on the high local merger rate, we find that 
over the course of a year, Advanced LIGO+ \cite{Aasi:2013wya} will typically have no events 
merging within the same second, and only a few occurrences of a BBH merging in the duration of a 
BNS (assuming $f_{\rm low} = 15$ Hz). For Voyager \cite{Adhikari_2020} we find 
$\mathcal{O}(1)$ instances of two events merging within the same second, and BNS signals 
will typically have at most one other signal's merger in their duration (for $f_{\rm low} = 10$ Hz). 
These numbers refer to signals detectable with a single interferometer (with SNR threshold 8) 
rather than with a network of them, but it will be clear that overlapping signals are going to 
become an important consideration mainly in the 3G era.

\begin{table}[t]
\renewcommand{\arraystretch}{1.2}
	\begin{tabular}{|c|c|c|c|}
		\hline
		Rate & Number of   & Number of & Number of \\
		    & BBH mergers & BNS mergers & any type  \\  
		\hline
		Low rate  & $8_{-5}^{+10}$ & $16_{-8}^{+16}$ & $25_{-12}^{+23}$ \\
		Median rate  & $13_{-7}^{+14}$ & $62_{-27}^{+58}$ & $76_{-33}^{+77}$  \\
		High rate   & $21_{-11}^{+21}$ & $157_{-66}^{+144}$ & $178_{-75}^{+164}$ \\
		\hline
	\end{tabular}
	\caption{Typical numbers of compact binary mergers happening during the time a BNS signal 
	is in band.}
	\label{tab:BNSover}
\end{table}

\section{Parameter estimation setup}
\label{sec:method}

Having established that third-generation detectors will see a considerable number of overlapping 
signals whose mergers occur very close in time, we want to find out what this will imply for 
parameter estimation. To this end, we simulate BBH and BNS signals in a network consisting of 
one ET and two CE observatories as in the previous section, assuming stationary, Gaussian noise following 
the PSDs used above. 

\begin{table}[t]
\renewcommand{\arraystretch}{1.2}
	\begin{tabular}{|c|c|c|c|}
		\hline
		Run & BBH-BBH   & BBH-BNS   & BNS-BNS \\
		\hline
		Low rate & $5$ & $57$ & $416$ \\
		Median rate & $11$ & $304$ & $6752$ \\
		High rate & $15$ & $1594$ & $41306$ \\
		\hline
	\end{tabular}
	\caption{Number of pairs of binary coalescence events with both SNRs between 15 and 30, and 
	such that their mergers occur within 2 seconds or less from each other.}
	\label{tab:2SecAnalysis}
\end{table} 

Since we expect parameter estimation biases to be more pronounced when SNRs of overlapping signals are similar to each
other, and on the low side, we focus on network SNRs roughly between 15 and 30. 
We consider overlapping events whose merger times either coincide (as a proxy for 
merger within the same second), or are separated by 2 seconds, again because these are the types of 
scenarios where biases will likely be the largest.  
The number of overlaps from the previous section that satisfy these criteria is given in Table~\ref{tab:2SecAnalysis}, for
different local merger rates; we see that they will be fairly common. 

\begin{figure}[ht!]
    \centering
    \includegraphics[width=0.48\textwidth]{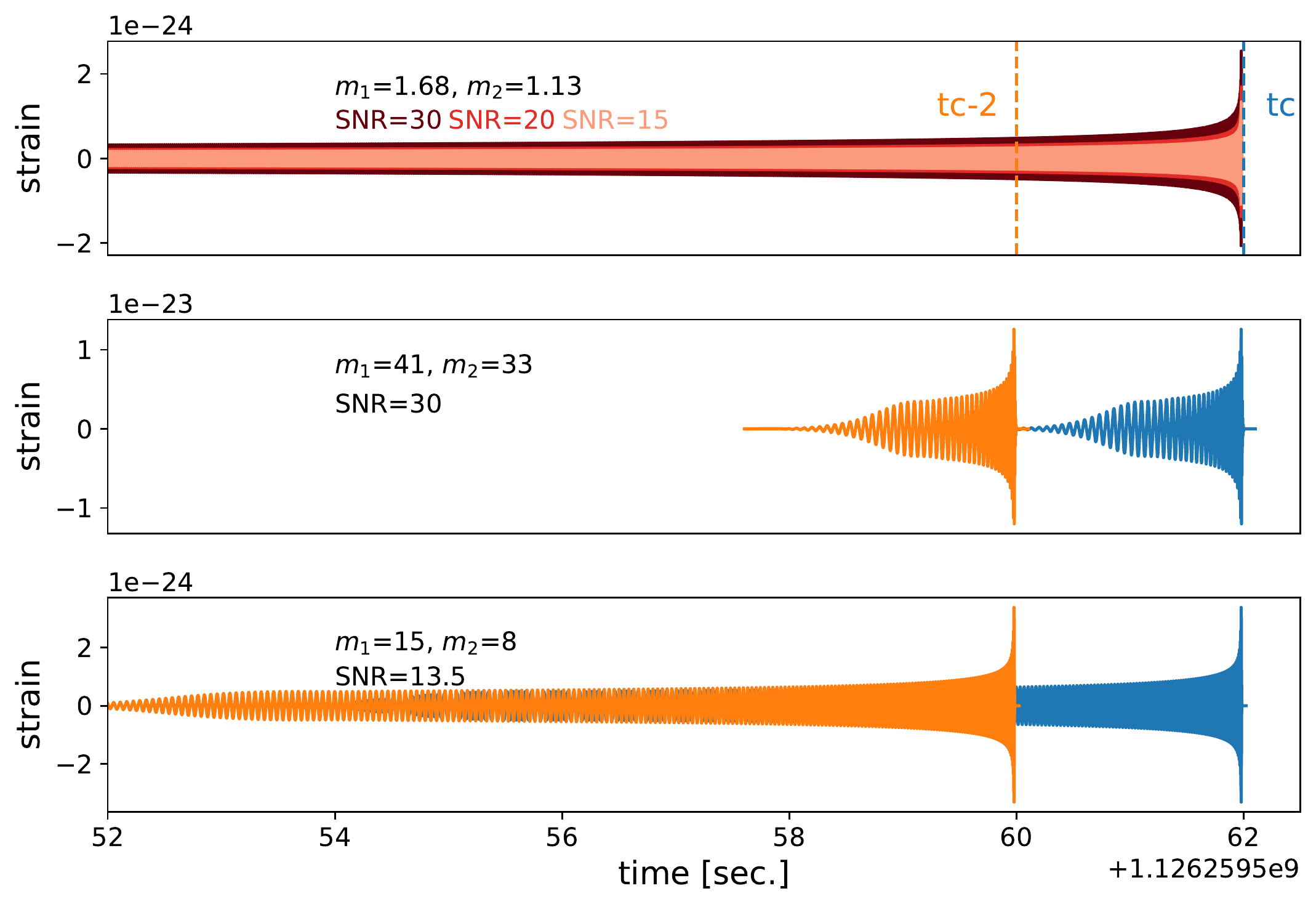}
    \vspace*{0.15cm}
    
    \includegraphics[width=0.48\textwidth]{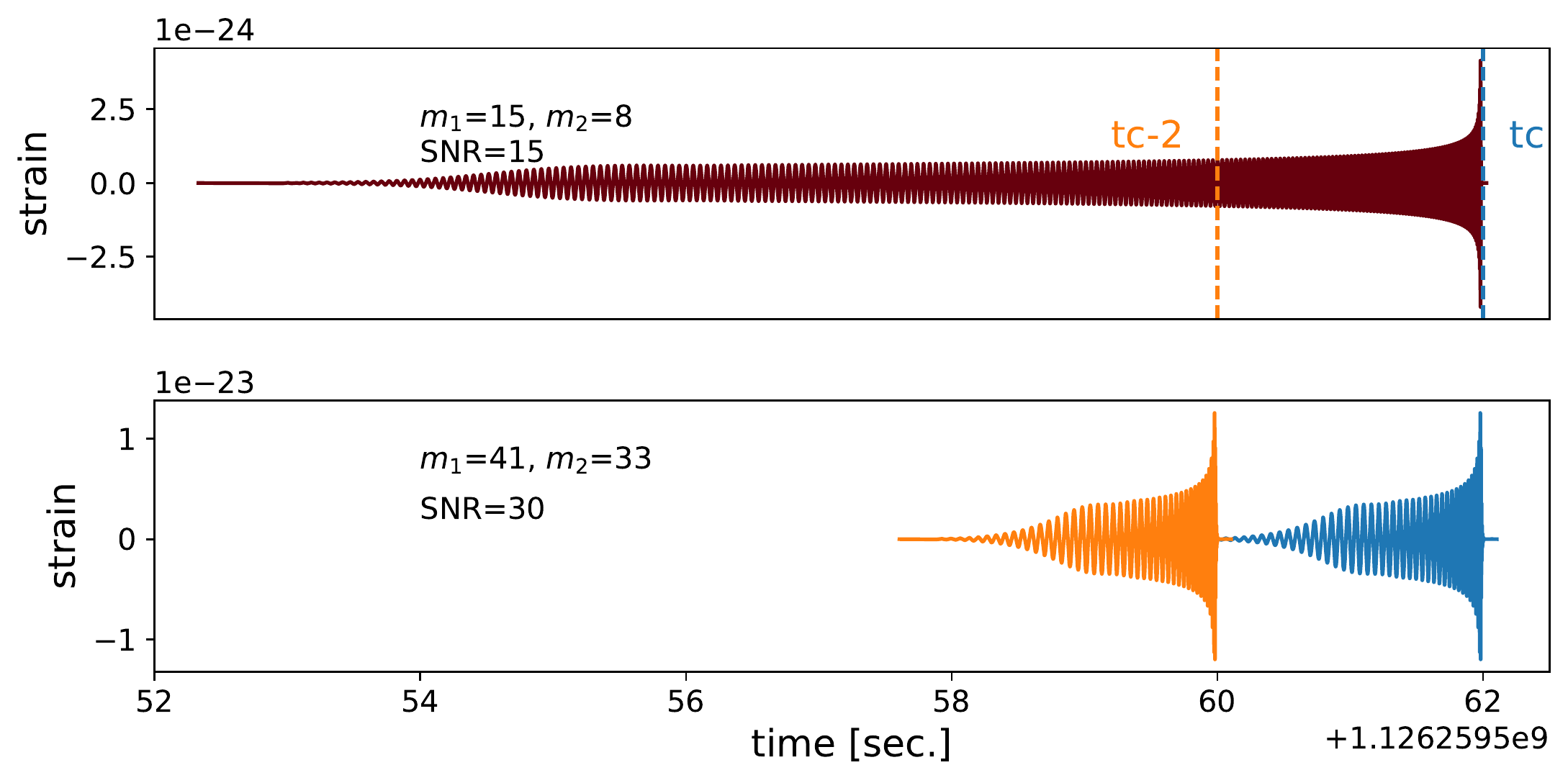}
    \vspace*{0.15cm}
    
    \includegraphics[width=0.48\textwidth]{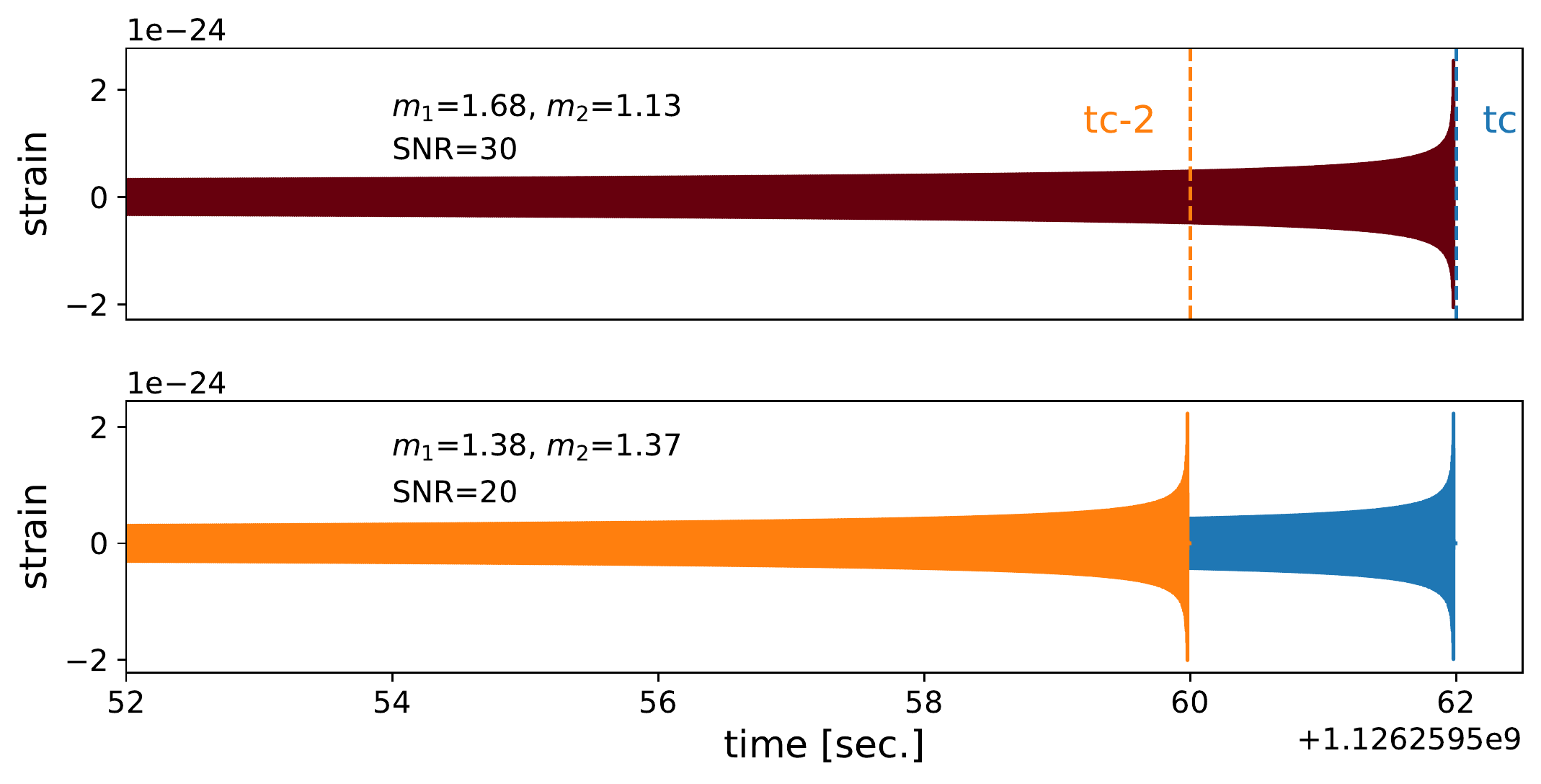}
    \caption{Individual waveforms and the overlap scenarios considered in our simulations.  
             All signals are injected in 3 different simulated noise realizations for a third-generation 
             detector network.
             Signals are either overlapped using the same end time (blue waveforms), or 2 seconds earlier 
             than the ``primary'' signal's end-time (orange waveforms). 
             \textit{Top three panels:} BNS signals (top) 
             with an SNR of 30, 20, or 15 being overlapped with 
             either a high-mass BBH signal (middle; GW150914-like) 
             or a low-mass BBH signal (bottom; GW151226-like).
             \textit{Middle panels:} Overlapping waveforms in the case of two BBH signals. 
             The higher-mass BBH signal (bottom; GW150914-like) 
             is overlapped with the lower-mass BBH signal (top; GW151226-like).
             \textit{Bottom panels:} Overlapping waveforms in the case of two BNS signals.
             }
    \label{fig:inj_all}
\end{figure}

In our parameter estimation studies, for definiteness we take the BBH events to have masses 
similar to those of either GW150914 (a higher-mass, shorter-duration signal) or GW151226 (a lower-mass, 
longer-duration event), while for BNSs we take the masses to be similar to those of GW170817. 
Overlapping signals are given different injected sky locations.  
All analyses are done with 3 different noise 
realizations. For each example of overlapping signals, parameter estimation is also done on the 
individual signals, for the same noise realizations, in order to assess what biases occur. 
Fig.~\ref{fig:inj_all} provides an overview of the various overlap scenarios that will be 
considered in the rest of this paper, in terms of masses and SNRs.

To reduce computational cost, we focus on non-spinning sources. A BBH signal is then 
characterized by parameters $\vec{\theta} = \{m_1, m_2, \alpha, \delta, \iota, \psi, D_{\rm L}, t_c, \varphi_c \}$,
where $m_1$, $m_2$ are the component masses, $(\alpha, \delta)$ specifies the sky position in terms of right
ascension and declination, $\iota$ and $\psi$ are respectively the inclination and polarization
angles which specify the orientation of the orbital plane with respect to the line of sight, 
$D_{\rm L}$ is the luminosity distance, and $t_c$ and $\varphi_c$ are respectively the time and 
phase at coalescence. BNS signals have two additional parameters $(\Lambda_1, \Lambda_2)$, 
corresponding to the (dimensionless) tidal deformabilities 
\cite{Flanagan:2007ix,PhysRevD.83.084051,Damour:2012yf,DelPozzo:2013ala,Agathos:2015uaa}.

In this work we focus specifically on potential biases in \emph{intrinsic} parameters. For BBHs, results
will be shown for the total mass $M = m_1 + m_2$ and mass ratio $q = m_2/m_1$ (with the 
convention $m_2 \leq m_1$). For BNSs, we show chirp mass $\mathcal{M}$ instead of total mass, since
that parameter is usually the best-determined one for long signals.
As the individual tidal deformabilities tend to be poorly measurable for the SNRs considered here, 
we will be showing results for a parameter $\tilde{\Lambda}$ defined as \cite{Wade:2014vqa}
\begin{equation}
\tilde{\Lambda}=\frac{16}{13} \sum_{i=1,2} \Lambda_i \frac{m_i^4}{M^4}\left( 12-11\frac{m_i}{M} \right), 
\end{equation} 
since this is how tidal deformabilities enter the waveform phase to leading (5PN) order 
\cite{Flanagan:2007ix}.  

In the Bayesian framework, all information about the parameters 
of interest is encoded in the posterior \emph{probability density function} (PDF), given 
by Bayes' theorem \cite{Veitch:2009hd}:
\begin{equation}
 p(\vec{\theta}|\mathcal{H}_s,d) 
 = \frac{p(d|\vec{\theta},\mathcal{H}_s)\,p(\vec{\theta}|\mathcal{H}_s)}{p(d|\mathcal{H}_s)},
 \label{eqn:Bayes}
\end{equation}
where $\vec{\theta}$ is the set of parameter values and $\mathcal{H}_s$ is the hypothesis that a 
GW signal depending on the parameters 
$\vec{\theta}$ is present in the data $d$. 
For parameter estimation purposes, the factor $p(d|\mathcal{H}_s)$, called the \emph{evidence} 
for the hypothesis  $\mathcal{H}_s$, is effectively set by the requirement that PDFs are normalized. 
Assuming the noise to be 
Gaussian, the \emph{likelihood} $p(d|\vec{\theta},\mathcal{H}_s)$ of obtaining data $d(t)$ given the 
presence of a signal $h(t)$ is determined by the proportionality
\begin{equation}
 p(d|\vec{\theta},\mathcal{H}_s) \propto \exp{\left [-\frac{1}{2}(d-h(\vec{\theta})|d-h(\vec{\theta}))\right ]},
 \label{eqn:lhood}
\end{equation}
where the noise-weighted inner product $(\,\cdot\,|\,\cdot\,)$ is defined as
\begin{equation}
(a|b) = 4\Re \int_{f_{\rm low}}^{f_{\rm high}} \frac{\tilde{a}^\ast(f)\,\tilde{b}(f)}{S_h(f)}\,df.
\end{equation}
Here a tilde refers to the Fourier transform, and $S_h(f)$ is the PSD, as in the previous section. 
Due to computational limitations, in our parameter estimation studies we use a lower frequency cut-off
of $f_{\mathrm{low}} = 23$ Hz. Since both ET and CE will be sensitive
down to lower frequencies than that, we expect that our choice will lead to conservative estimates of 
parameter estimation biases, as the same signal will in reality accumulate more SNR when it is visible
in the detector already from a lower frequency.  

Our choices for the \emph{prior probability density} $p(\vec{\theta}|\mathcal{H}_s)$ in Eq.~(\ref{eqn:Bayes}) 
are similar to what has been used for the analyses of real data when BBH or BNS signals were 
present with masses similar to the ones specified in Fig.~\ref{fig:inj_all}. In all cases we sample 
uniformly in component masses. For the GW150914-like signals, we do this in the range  
$m_1, m_2 \in [10, 80]\,M_\odot$.  
For analyzing the GW151226-like signals, the component mass range is $m_1, m_2 \in [3, 54.4]\,M_\odot$, 
and in addition we restrict chirp mass to $\mathcal{M} \in [5, 20]\,M_\odot$ and mass ratio 
$q$ to the range $[0.05, 1]$. For BNSs we sample component masses 
in the range $m_1, m_2 \in [1, 2]\,M_\odot$, restricting $\mathcal{M} \in [0.7,2]\,M_\odot$, while  
tidal deformabilities are sampled uniformly in the range $\Lambda_1, \Lambda_2 \in [0, 5000]$. 
When we show PDFs for the derived quantity $\tilde{\Lambda}$, they will have been reweighted with the 
prior probability distribution of this parameter induced by the flat priors 
on component masses and $\Lambda_1$, $\Lambda_2$, such as to effectively have a uniform
prior on $\tilde{\Lambda}$.

To sample the likelihood function in Eq.~(\ref{eqn:lhood}), we use the LALInference library 
\cite{Veitch:2014wba}, and specifically the \texttt{lalinference\_mcmc} algorithm. 
The waveforms we use for the BNS and BBH signals are \texttt{IMRPhenomD\_NRTidalv2} 
\cite{Dietrich:2017aum,Dietrich:2018uni,Dietrich:2019kaq} and 
\texttt{IMRPhenomD} \cite{Husa:2015iqa,PhysRevD.93.044007} respectively, both computed with the waveform 
library LALSimulation. 
To inject the signals and add noise to them, we use standard tools available within the 
LALSimulation package. All these codes are openly accessible in LALSuite \cite{lalsuite}. 

Before performing parameter estimation, we verify the detectability of the individual signals in the
overlap scenarios of Fig.~\ref{fig:inj_all} using the PyCBC software package  
\cite{Biwer:2018osg}. We inject overlapping signals in noise generated from the PSD and check that the 
individual signals show up as triggers with masses that are consistent between detectors, at a 
network SNR above a threshold of 8. This turns out to be true for all the cases considered, except for 
two BBH signals merging at the same time. In the latter case we still have triggers in individual 
detectors, but with masses differing by up to $\sim 5 M_{\odot}$. Using the SNRs in single 
detectors as detection statistics, detection is still achieved. For all scenarios, the end 
times of individual signals tend to be identified with a precision of a few milliseconds  
\cite{VanDenBroeck:2006ar}; when subsequently performing parameter estimation, we use a prior 
range for end time that is centered on the true end time, leaving an interval of 0.1 s on either side. 

For parameter estimation, all simulations are done with three different noise realizations. 
In the next section, results are shown for one of those; for the other two noise realizations, 
see Appendix \ref{sec:appendix}. 

As usual, the one-dimensional PDF $p(\lambda | \mathcal{H}_s,d)$ for a particular parameter $\lambda$ is obtained 
from the joint PDF $p(\vec{\theta}|\mathcal{H}_s,d)$ by integrating out all other parameters. In 
assessing the effect on parameter estimation of signals that overlap in various ways, we will frequently
be comparing one-dimensional PDFs for the same parameter in different situations. A convenient
way of quantifying the difference between two distributions $p_1(\lambda)$ and $p_2(\lambda)$ is
by means of the \emph{Kolmogorov-Smirnov (KS) statistic} \cite{Kolmogorov,Smirnov}. Let $P_1(\lambda)$, 
$P_2(\lambda)$ be the associated \emph{cumulative} distributions; then the KS statistic is just 
the largest distance between these two:
\begin{equation}
\mbox{KS} = \mbox{sup}_\lambda | P_1(\lambda) - P_2(\lambda) |.
\label{KS}
\end{equation}
By construction, this yields a number between 0 and 1; if the KS statistic is close to zero, then 
the distributions $p_1(\lambda)$ and $p_2(\lambda)$ will be considered close to each other.

\section{Results}
\label{sec:results}
\begin{figure}[t!]
    \centering
    \includegraphics[width=0.5\textwidth]{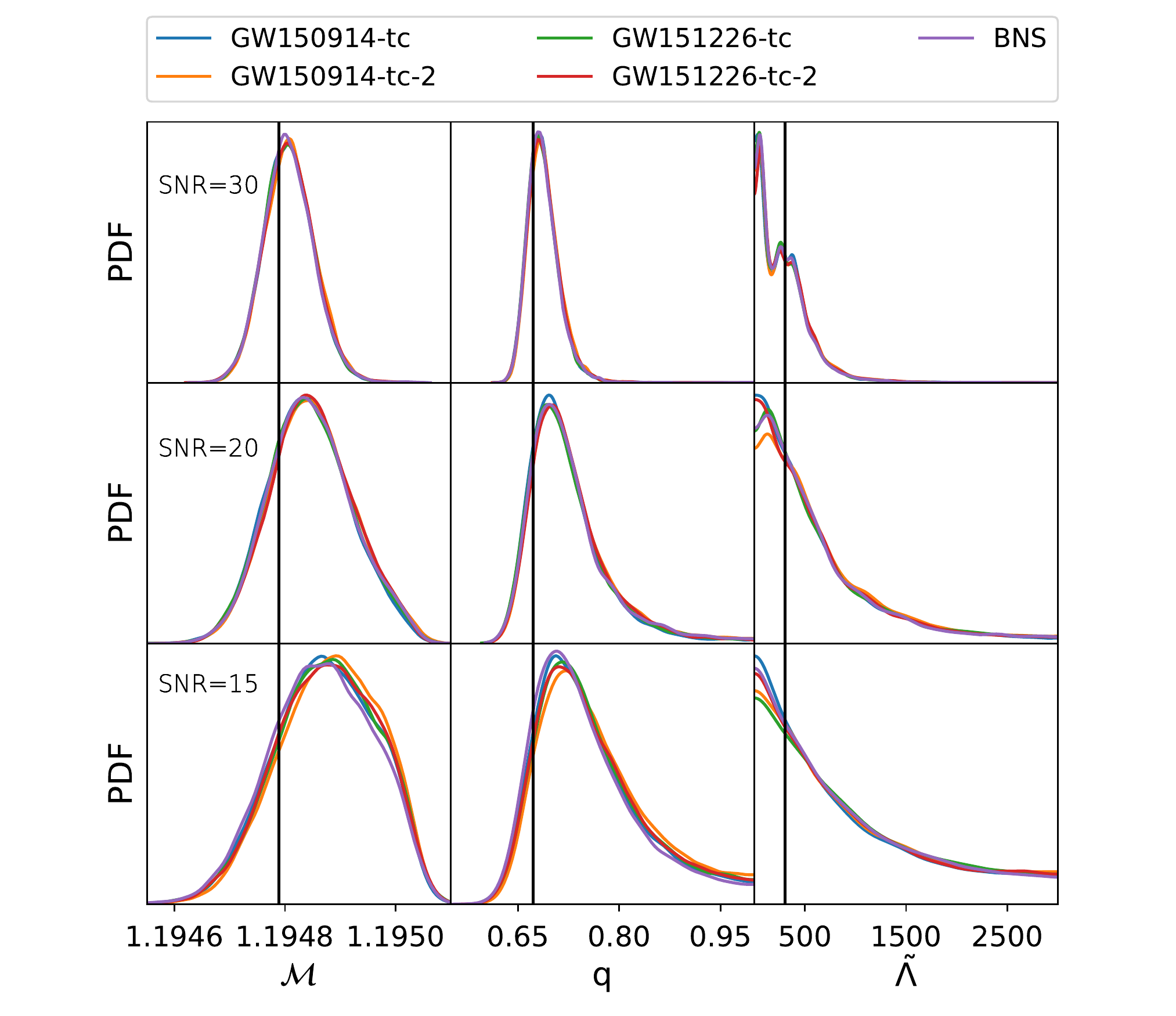}
    \caption{Posterior PDFs showing estimation of intrinsic parameters when the BNS signal has SNR = 30 
    (top row), SNR = 20 (middle row), and SNR = 15 (bottom row).  
             Results are shown for the cases 
             when the GW150914-like signal ends at the same time 
             as the BNS signal (\texttt{GW150914-tc}), when it ends 2 seconds earlier 
             (\texttt{GW150914-tc-2}), when the GW151226-like signal ends at the 
             same time as the BNS (\texttt{GW151226-tc}), when it ends 
             2 seconds earlier 
             (\texttt{GW151226-tc-2}), and finally when the injected 
             signal is only the BNS (\texttt{BNS}). The true values of the 
             parameters are indicated by vertical, black lines.
            }
    \label{fig:intrinsic_overlap_1}
\end{figure}

\begin{table*}[t]
    \begin{tabular}{| *{10}{c|} }
    \hline
BBH overlapped    & \multicolumn{3}{c|}{BNS (SNR = 30)}
                    & \multicolumn{3}{c|}{BNS (SNR = 20)}
                        & \multicolumn{3}{c|}{BNS (SNR = 15)}  \\
    \hline
   &   $\mathcal{M}$  &   $q$  &   $\tilde{\Lambda}$  &   $\mathcal{M}$  &   $q$   &   $\tilde{\Lambda}$  &   $\mathcal{M}$   &   $q$ & $\tilde{\Lambda}$  \\
   \hline
   \texttt{GW150914-tc} & 0.0112 & 0.00915 & 0.0277 & 0.0162 & 0.0204 & 0.0275 & 0.0297 & 0.0323 & 0.00947 \\
   \hline
   \texttt{GW150914-tc-2} & 0.0320 & 0.0389 & 0.0168 & 0.0235 & 0.0273 & 0.0331 & 0.0704 & 0.0840 & 0.0218 \\
   \hline
   \texttt{GW151226-tc} & 0.00754 & 0.00748 & 0.0113 & 0.0123 & 0.0139 & 0.0173 & 0.0403 & 0.0516 & 0.0305 \\
   \hline
   \texttt{GW151226-tc-2} & 0.0187 & 0.0220 & 0.0309 & 0.0227 & 0.0233 & 0.0259 & 0.0521 & 0.0513 & 0.0159 \\
    \hline

    \end{tabular}
    \caption{Values of the KS statistic comparing PDFs for BNS parameters (columns) in the BNS+BBH overlap scenarios (rows) with the 
    corresponding PDFs when there is no overlapping BBH signal. The small numbers indicate absence of significant bias.
    The numbers shown here correspond to the PDFs in Fig.~\ref{fig:intrinsic_overlap_1}.}
    \label{tab:KS_BBH_BNS_noise2}
\end{table*}

\begin{figure}[t!]
    \centering
    \includegraphics[width=0.5\textwidth]{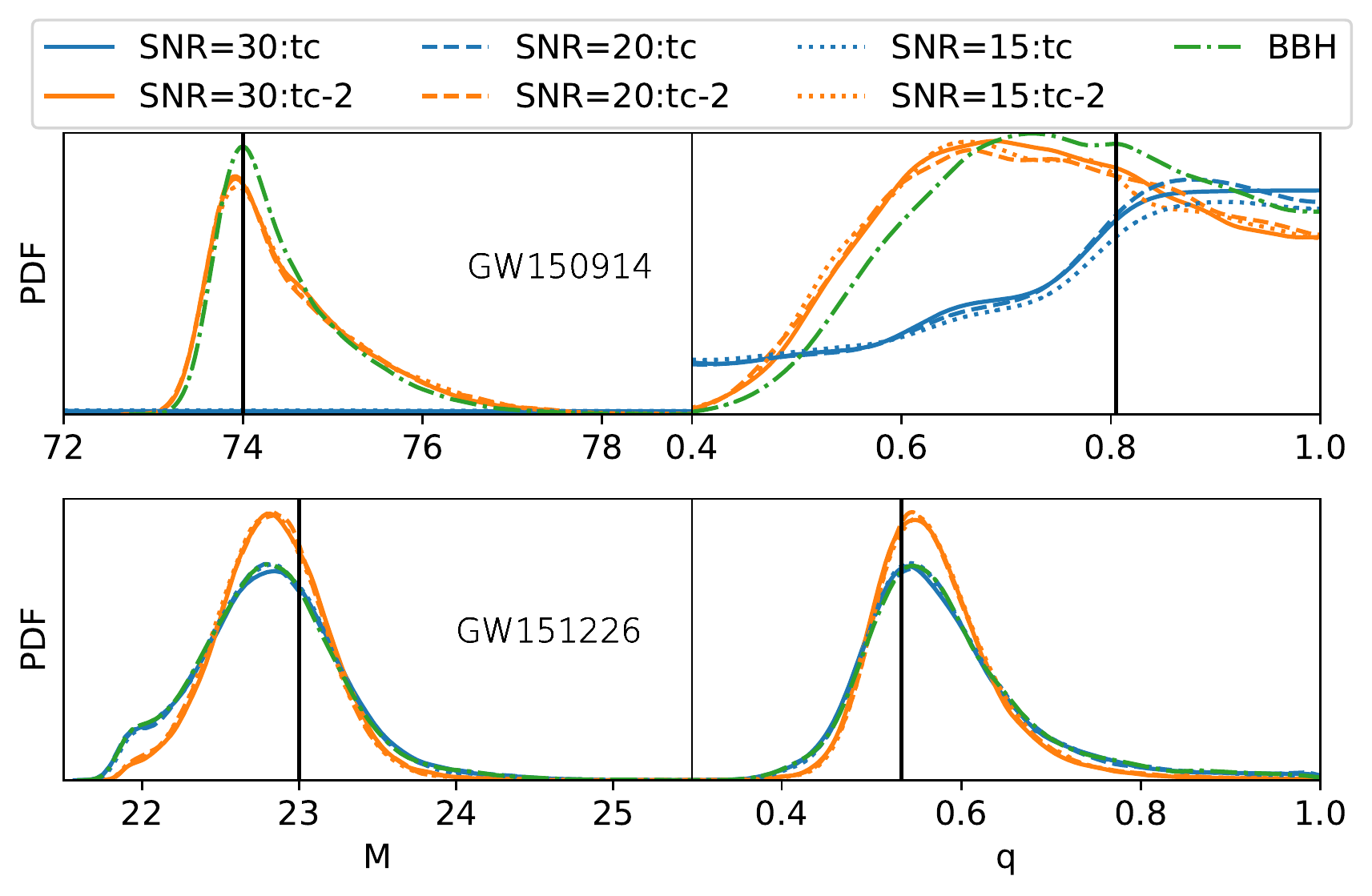}
    \caption{Posterior PDFs for total mass and mass ratio, for the GW150914-like signal 
    (top panel) and the GW151226-like signal (bottom panel) when they are respectively being overlapped with 
    a BNS signal of SNR = 30 (solid lines), SNR = 20 (dashed lines), and  
    SNR = 15 (dotted lines). The overlaps are being done when 
    the BBH and the BNS end at the same time (\texttt{tc}), and when the  
    BBH ends 2 seconds before the BNS (\texttt{tc-2}). 
    Finally, posterior PDFs for the two BBH signals by themselves are shown as 
    green, dashed-dotted lines (\texttt{BBH}). 
    The injected parameter values are indicated by black,  
    vertical lines.}
    \label{fig:intrinsic_overlap_12}
\end{figure}

\begin{table*}[t]
    \begin{tabular}{| *{9}{c|} }
    \hline
BNS overlapped   & \multicolumn{2}{c|}{\texttt{GW150914-tc}}
                    & \multicolumn{2}{c|}{\texttt{GW150914-tc-2}}
                        & \multicolumn{2}{c|}{\texttt{GW151226-tc}}
                           & \multicolumn{2}{c|}{\texttt{GW151226-tc-2}} \\
    \hline
   &   $M$   &   $q$  &   $M$   &   $q$  &   $M$  &   $q$  &   $M$  &   $q$  \\
   \hline
   BNS (SNR = 15) & -- & -- & 0.0504 & 0.0807 & 0.00933 & 0.0117 & 0.0687 & 0.0657 \\
   \hline
   BNS (SNR = 20) & -- & -- & 0.0427 & 0.0698 & 0.0107 & 0.0106 & 0.0727 & 0.0700 \\
   \hline
   BNS (SNR = 30) & -- & -- & 0.0379 & 0.0673 & 0.0187 & 0.183 & 0.0819 & 0.0793 \\
   \hline
    \end{tabular}
    \caption{Values of the KS statistic comparing PDFs for BBH parameters (columns) in the BNS+BBH overlap scenarios  
    (rows) with the corresponding PDFs when there is no overlapping BNS signal. In the case of a GW150914-like signal merging at the same 
    time as a BNS, the sampler fails to find the signal, but other scenarios are not so problematic. 
    For GW151226, the slightly higher values for the \texttt{tc-2} case compared to the \texttt{tc} case are likely due to
    the signals being placed in a slightly different part of the noise stream (two seconds earlier) from the BBH-only cases
    that are used for comparison.
    The numbers shown here correspond to the PDFs in Fig.~\ref{fig:intrinsic_overlap_12}.
    }
    \label{tab:KS_BBHRec_BBH_BNS_noise2}
\end{table*}

\subsection{Overlap of a BNS signal with a BBH signal}
\label{subsec:BNSwithBBH}
First we look at the results of parameter estimation for the overlap of a BNS signal with a BBH, either 
ending at the same time or with the BBH signal ending 2 seconds earlier than the BNS. 
This is the scenario shown in the top panels of Fig.~\ref{fig:inj_all}. We perform parameter 
estimation first on the BNS and then on the BBH, with priors as specified in the previous section.

\subsubsection{BNS recovery}
Fig.~\ref{fig:intrinsic_overlap_1} shows posterior probability distributions for 
intrinsic parameters characterizing the BNS signal, for 3 different SNRs of the BNS, and 
the different overlap scenarios. The PDFs tend to widen with 
decreasing SNR, as expected. We see that estimation of the mass parameters are essentially unaffected, 
regardless of the type of overlapping BBH signal (GW150914-like or GW151226-like) or
of its merger time (identical to that of the BNS, or 2 seconds earlier). For a given SNR
of the BNS, the PDFs for the tidal parameter $\tilde{\Lambda}$ differ slightly more between the overlap
scenarios. However, we note that most of the information on tides enters the signal at high frequencies,
where the detectors are less sensitive; and in fact, as shown in Appendix \ref{sec:appendix} 
(Fig.~\ref{fig:noise23_rec_bns_bbh_plus_bns_inj}), differences in the underlying noise
realization tend to have a larger effect on the measurement of $\tilde{\Lambda}$ than overlapping
signals. 

We conclude that an overlapping BBH signal does not have much impact on the estimation of the BNS
parameters, even if the BBH merger time is arbitrarily close to that of the BNS. This is corroborated
by the KS statistics in Table \ref{tab:KS_BBH_BNS_noise2}, which compare PDFs for the various 
overlap scenarios with the corresponding PDFs in the absence of overlapping signals. It is 
reasonable to assume that placing a BBH signal even earlier in the BNS would also 
have had little impact.

\subsubsection{BBH recovery}
Figure~\ref{fig:intrinsic_overlap_12} shows parameter estimation on the BBHs 
when the SNR of the BNS signal is varied from 30, to 20, to 15. Table \ref{tab:KS_BBHRec_BBH_BNS_noise2} has
the corresponding KS statistics comparing with PDFs obtained in the absence of overlap. Again results are shown 
for a particular noise realization; see Fig.~\ref{fig:noise23_rec_bbh_bbh_plus_bns_inj} 
in Appendix \ref{sec:appendix} for two other noise realizations. 
We see that when the BBH signal has a time of coalescence 2 seconds earlier than the BNS 
(\texttt{tc-2} in the figure), the signal 
is well recovered. However, when the BBH signal and the BNS signal end at the same instant of time, 
the BBH recovery deteriorates, and in the case of the GW150914-like signal, 
the sampling process in fact fails to find the signal.  
For the GW151226-like signal, while the estimates are offset from their true values, there is some 
measurability of the signal when the times of coalescence of the BBH and BNS are the same. 
The different outcomes between the GW150914-like and GW151226-like injection are likely due to  
the short duration of the GW150914-like signal, effectively leading to a distortion of the entire signal 
when the merger happens at the same instant as the BNS merger. By contrast, the much longer inspiral 
of the GW151226-like signal implies many more wave cycles for the parameter estimation algorithm to latch 
on to. Finally, as the SNR over the underlying BNS signal is varied (keeping the SNR of the BBH signal 
the same), the PDFs for the BBH show essentially no change.  
Placing a BBH signal only 2 seconds before the BNS merger causes the BBH to be recovered 
without appreciable biases, so it is reasonable to assume that placing a BBH signal still earlier in the 
BNS inspiral would also have little effect on its recovery.

\subsection{Overlap of 2 BBH signals}

\begin{figure}[t!]
    \centering
    \includegraphics[width=0.5\textwidth]{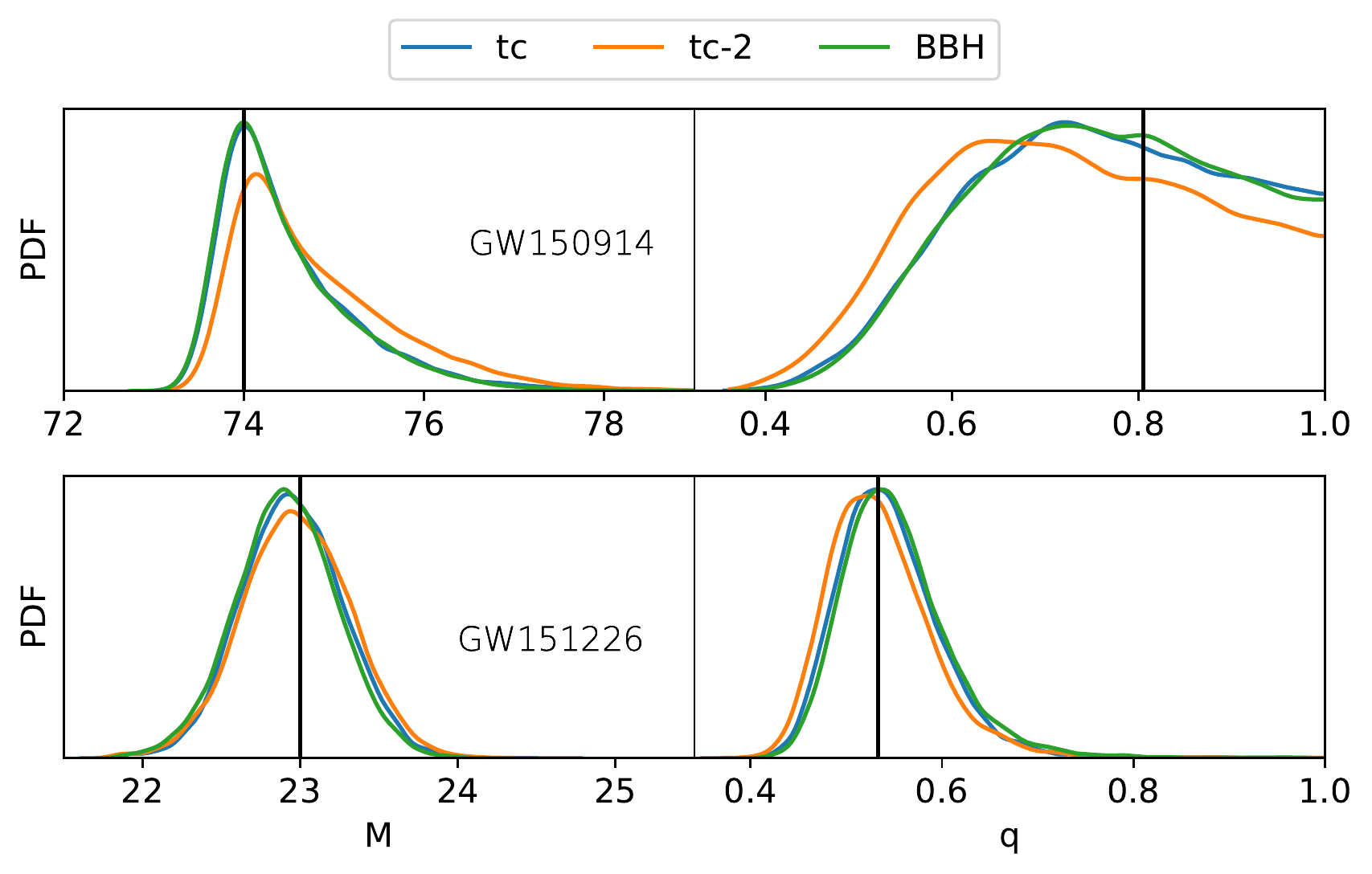}
    \caption{Posterior PDFs for total mass and mass ratio when a GW150914-like signal and 
    a GW151226-like signal 
    are being overlapped at the same trigger times (\texttt{tc}) and when the trigger time of the 
    GW150914-like BBH ends 2 seconds earlier (\texttt{tc-2}), compared with 
    parameter estimation in the absence of overlap (\texttt{BBH}). The top panel shows the recovery of 
    the GW150914-like signal and the bottom one that of the GW151226-like signal. Black vertical 
    lines indicate the true values of the parameters.}
    \label{fig:2BBHs_noise}
\end{figure}

\begin{table*}[t]
    \begin{tabular}{| *{9}{c|} }
    \hline
    \multicolumn{2}{|c|}{\texttt{GW150914-tc}}
                    & \multicolumn{2}{c|}{\texttt{GW150914-tc-2}}
                        & \multicolumn{2}{c|}{\texttt{GW151226-tc}}
                           & \multicolumn{2}{c|}{\texttt{GW151226-tc-2}} \\
    \hline
    $M$  &   $q$  &   $M$   &   $q$  &   $M$  &   $q$  &   $M$  &   $q$  \\
   \hline

   0.0195 & 0.0109 & 0.162 & 0.103 & 0.0446 & 0.0478 & 0.0844 & 0.127  \\
   \hline
    \end{tabular}
    \caption{Values of the KS statistic comparing PDFs for BBH parameters in the BBH+BBH overlap scenarios  
    with the corresponding PDFs without an overlapping signal. 
    The slightly higher values for the \texttt{tc-2} cases are likely due to
    the signals being in a slightly different part of the noise stream (two seconds earlier) from the BBH-only cases
    used for comparison.
    However, in all cases there is no significant bias. 
    The numbers shown here correspond to the PDFs in Fig.~\ref{fig:2BBHs_noise}.}
    \label{tab2:BBHs_noise_KS}
\end{table*}

The scenario being analyzed here is the one in the middle panels of Fig.~\ref{fig:inj_all}. 
Fig.~\ref{fig:2BBHs_noise} shows the posterior PDFs on total mass $M$ and mass ratio $q$ 
when two BBH signals of different masses are being overlapped, compared with parameter 
estimation on the same signals in situations where there is no overlap (\texttt{BBH}). 
The corresponding KS statistic values are given in Table \ref{tab2:BBHs_noise_KS}.  
We find the results to be consistent within 
statistical fluctuations. Here too, the signals are overlapped once with the same 
coalescence times (\texttt{tc}), and once with one of the signals, GW150914, ending 2 seconds earlier 
(\texttt{tc-2}). The SNRs of the two signals, GW150914-like, and GW151226-like, are 30 and 15, 
respectively. As can be seen in the Figure, the two BBH signals' parameters can be extracted 
without any biases even when they end simultaneously. Again see Appendix \ref{sec:appendix}
for other noise realizations, with the same conclusion.

\subsection{Overlap of 2 BNS signals}

\begin{figure}[t!]
    \centering
    \includegraphics[width=0.5\textwidth]{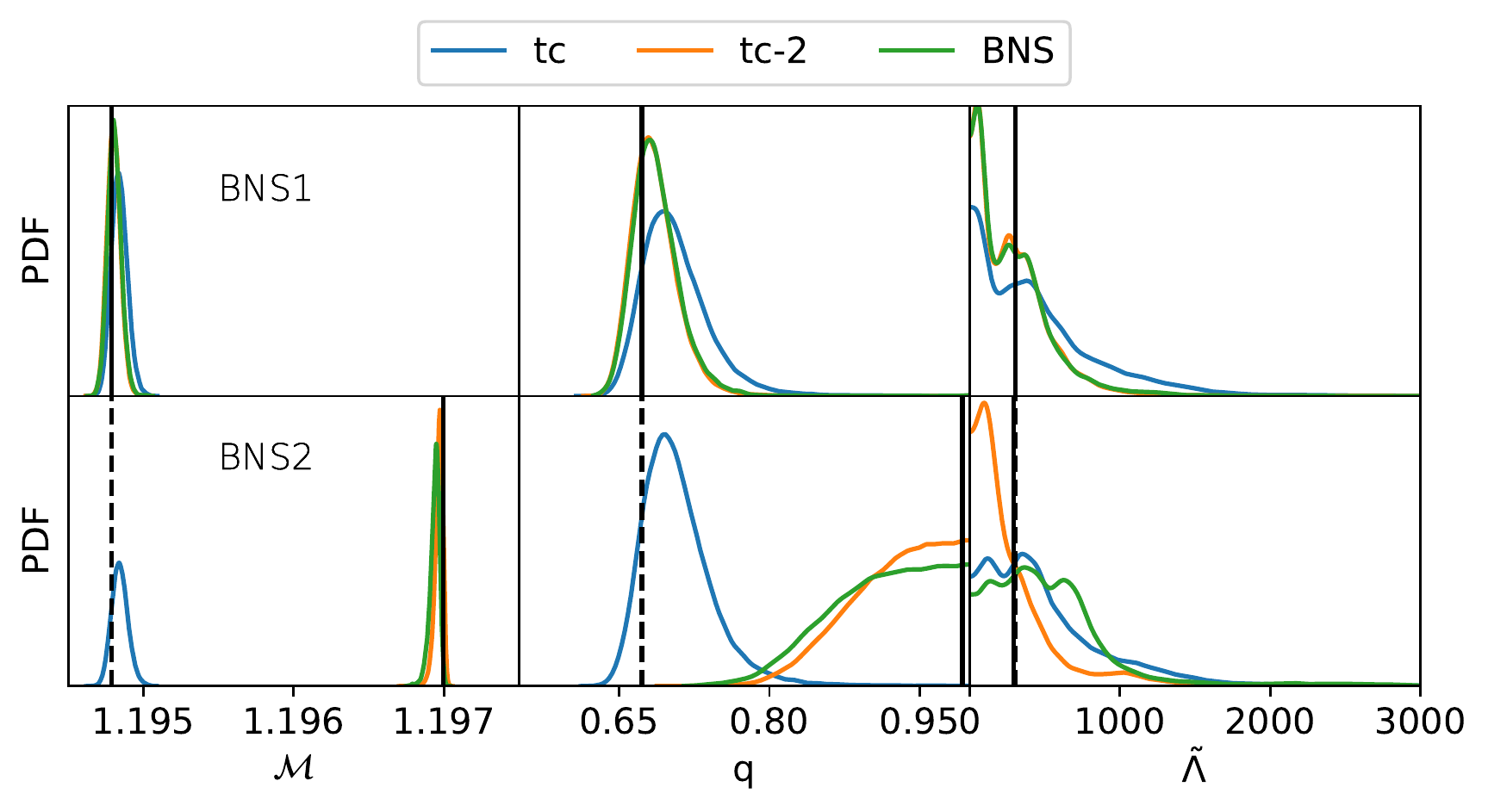}
    \caption{Posterior PDFs showing recovery on chirp mass, mass ratio and tidal deformability 
    $\tilde{\Lambda}$ when two BNSs, referred to as BNS1 and BNS2, are being overlapped at the same 
    time of coalescence (\texttt{tc}) and when BNS2 ends 2 seconds earlier than 
    BNS1 (\texttt{tc-2}). These are compared
    with results in the absence of overlap (\texttt{BNS}). The top panel is for the 
    recovery of BNS1 and the bottom one for the recovery of BNS2. 
    The solid black vertical lines indicate the injected values of the source being recovered each time. 
    We note that when the times of coalescence of the two BNSs are the same, the parameter 
    estimates recovered are those of BNS1, whose injected values are 
    also shown in the bottom panel, as dashed vertical black lines. 
    }
    \label{fig:2BNSs_noise}
\end{figure}

\begin{table*}[t]
    \begin{tabular}{| *{12}{c|} }
    \hline
\multicolumn{3}{|c|}{BNS1 (\texttt{tc})}
                      & \multicolumn{3}{c|}{BNS1 (\texttt{tc-2})}
                        & \multicolumn{3}{c|}{BNS2 (\texttt{tc})}  
                            & \multicolumn{3}{c|}{BNS2 (\texttt{tc-2})}\\
    \hline
      $\mathcal{M}$  &   $q$  &   $\tilde{\Lambda}$  &   $\mathcal{M}$ &  $q$  &   $\tilde{\Lambda}$  &  $\mathcal{M}$   &  $q$  & $\tilde{\Lambda}$ &   $\mathcal{M}$  & $q$  & $\tilde{\Lambda}$  \\
  
   \hline
    0.269 & 0.270 & 0.202 & 0.0309 & 0.0216 & 0.0129 & 1.0 & 0.955 & 0.0762 & 0.384 & 0.0951 & 0.368 \\
   \hline

    \end{tabular}
    \caption{Values of the KS statistic comparing PDFs for BNS parameters in the BNS+BNS overlap scenarios  
    with the corresponding PDFs without an overlapping signal; see also Fig.~\ref{fig:2BNSs_noise}. 
    We see that the numbers are higher for both BNSs when they end at the same time; in fact, the measured
    parameters for BNS2 are those of BNS1.  However, when the BNSs merge 2 seconds apart, the values are
    much lower, showing that the biases largely disappear. 
    The numbers shown here correspond to the PDFs in Fig.~\ref{fig:2BNSs_noise}.}
    \label{tab:2BNS_noise_KS}
\end{table*}

Finally, we analyze the simulations in the bottom panels of Fig.~\ref{fig:inj_all}. 
Figure~\ref{fig:2BNSs_noise} shows the recovery of BNS parameters for each BNS signal when two 
BNS signals are being overlapped, again with either the same  
coalescence times and when one of the BNSs (henceforth BNS2) ends 2 seconds earlier than the other 
BNS signal (henceforth BNS1). For KS statistic values comparing PDFs with the corresponding non-overlapping cases, 
see Table \ref{tab:2BNS_noise_KS}.
BNS1 and BNS2 respectively have SNRs of 30 and 20, and 
component masses $(m_1, m_2) = (1.68, 1.13)\,M_\odot$ and $(m_1, m_2) = (1.38, 1.37)\,M_\odot$.  
These particular choices cause both signals to have very similar chirp masses. 
Given these masses, their tidal deformabilities, $\tilde{\Lambda}=303$ for BNS1 and 
$\tilde{\Lambda}=292$ for BNS2,  
follow the equation of state APR4; these were the simulated signals used 
for investigating systematics in the measurements on GW170817 in Ref.~\cite{Abbott:2018wiz}. 

In Fig.~\ref{fig:2BNSs_noise}, the top panel shows the posterior PDFs on chirp mass, mass ratio, 
and tidal deformability for  
BNS1 when BNS2 ends at the same time (\texttt{tc}) and when BNS2 ends 2 seconds earlier 
(\texttt{tc-2}), together with the case where only BNS1 is present in the data (\texttt{BNS}). 
The bottom panels show the same, but for the recovery of BNS2. 
When the two signals end at the same time, the parameters 
characterizing BNS1 are being recovered, which likely happens because of the higher SNR of BNS1. 
As the tidal deformabilities of the two sources are so 
close, the PDFs for $\tilde{\Lambda}$ look similar in all cases.
However, also looking at the mass parameters, 
parameter estimation is rather robust when the signals end 2 seconds apart.

\section{Conclusions}
\label{sec:conclusion}

Given regular improvements in the sensitivity of gravitational-wave detectors and especially 
the planned construction of the next generation of interferometers, it will become increasingly 
likely that individually detectable gravitational-wave signals will end up overlapping in the data. 
In this paper we 
(i) assessed how often different types of overlap will happen in ET and CE, and 
(ii) tried to quantify the impact this would have on parameter estimation with current  
data analysis techniques. 

To address the question of the nature and frequency of different overlap scenarios, 
for each of three possible local merger rates, we constructed a 
``catalog'' of signals in ET and CE, enabling a more in-depth study of overlaps than in previous works. 
We showed that there will be a significant number of signals for which the merger happens within
the same second, varying from tens to thousands depending on the local merger rate. Additionally, 
the substantial increase in the duration of BNS events due to the improved low frequency sensitivity of
third generation observatories will lead to the occurrence of up to tens of other signals 
overlapping with a given BNS.

Motivated by these results, we performed the first detailed Bayesian analysis study on possible 
biases that may arise in future as detection rates become higher and overlapping signals start to occur. 
We focused on overlapping signals for which the end times were close to each other, so that in 
particular there is overlap at times where both signal amplitudes are high; it is this type of 
situation where we expect parameter estimation biases to be the most pronounced. Specifically, 
merger times were taken to be either the same (as a proxy for being arbitrarily close to each other),  
or separated by 2 seconds. 
Our preliminary conclusions (based on a limited number of investigations) are as follows:
\begin{itemize}
\item When BBH signals are overlapping with a BNS signal of similar SNR, parameter estimation
on the BNS is hardly affected, even with the merger time of the BBH arbitrarily close to that of the
BNS. Presumably this is due to the much larger number of BNS wave cycles in band compared to the BBH.
\item However, in the same scenario, parameter estimation on the BBH can be subject to significant
biases if the BBH is high-mass, so that its signal is short. That said, the problem largely disappears
when the BNS and BBH merger times are separated by 2 seconds, or when the BBH is low-mass.
\item When two BBHs with sufficiently dissimilar masses overlap with close-by merger times, 
parameter estimation on either of the signals will not be much affected.
\item When two BNS signals overlap with close-by merger times, parameter estimation will recover 
the louder signal reasonably well. With a 2 second separation of merger times, good-quality 
parameter estimation can already be done on the two signals separately. 
\end{itemize}

These results suggest that current parameter estimation techniques will, in several types of  
situations of interest, already perform reasonably well in the 3G era when applied to overlapping signals, 
even when the individual signals have similar SNRs, and even when the SNRs are on the low side given 
the projected distribution for these observatories. Nevertheless, a number of questions remain. 
What happens when SNRs are gradually increased? Related to this is the choice of lower cut-off frequency; 
to what extent will parameter estimation improve as one goes to $f_{\rm low} = 5$ Hz or even lower, 
so that signals have a much larger number of wave cycles in the detector's sensitive band? 
Though not the focus here, at higher SNRs the use of currently available waveform approximants to analyze BNS signals 
in 3G detectors would lead to biases in the estimation of $\tilde{\Lambda}$ even 
in the absence of overlap \cite{Samajdar:2018dcx}, also motivating further research in waveform modeling.
Spins were not included in our study, but it would be of interest to see their effect: 
large precessing spins will complicate parameter estimation in the case of BBHs, while for BNSs, 
having access to the spin-induced quadrupole moment can aid in determining tidal deformabilities 
\cite{Samajdar:2019ulq}. Finally, what happens when overlaps involve (much) more than two signals, 
e.g.~a long BNS signal overlapping with a large number of BBH signals? These questions are left 
for future work.

In order to make optimal scientific use of the capabilities of 3G detectors, it will be appropriate to 
develop Bayesian parameter estimation techniques for which the likelihood function assumes multiple 
signals to be present in a given stretch of data, e.g.~replacing Eq.~(\ref{eqn:lhood}) by
\begin{eqnarray}
&& p(d|\{\vec{\theta}_i\},\mathcal{H}_s) \nonumber\\
&&\propto \exp{\left[-\frac{1}{2}\bigg(d-\sum_{i = 1}^N h(\vec{\theta_i})\bigg|d-\sum_{i=1}^N h(\vec{\theta}_i)\bigg)\right]},
 \label{eqn:extlhood}
\end{eqnarray}
with $N$ the number of signals found by a detection pipeline, and $\vec\theta_i$, $i = 1, \ldots, N$ the 
associated parameters. Additionally, one could 
let $N$ itself be a parameter to be sampled over, thus allowing for an a priori unknown number
of signals in the given stretch of data. In all this, it may be possible to borrow from techniques developed 
in the context of somewhat related problems in GW data analysis, such as the characterization 
of the large number of (in this case near-monochromatic) signals from galactic white dwarf binaries in 
the space-based LISA~\cite{Crowder:2004ca,Cornish:2005qw,Cornish:2006ms,Crowder:2006eu,
Littenberg:2011zg,Robson:2017ayy,Littenberg:2020bxy}, 
BNSs in BBO~\cite{Cutler_2006}, or supermassive black hole binaries in pulsar timing searches
\cite{Petiteau:2012zq}.

\begin{acknowledgments} 
  We are grateful to Elia Pizzati, Surabhi Sachdev, Anuradha Gupta, and Bangalore Sathyaprakash 
  for sharing and discussing their results on a related study. 
  A.S., J.J, and C.V.D.B are supported by the research programme 
  of the Netherlands Organisation for Scientific Research (NWO).
  The authors are grateful for computational resources provided by the 
  LIGO Laboratory and supported by the National Science Foundation Grants No. PHY-0757058 and No. PHY-0823459. 
  We are grateful for computational resources provided by Cardiff University, and
  funded by an STFC grant supporting UK Involvement in the Operation of Advanced
  LIGO.
\end{acknowledgments}

\bibliography{refs}

\onecolumngrid

\appendix
\section{Parameter estimation for different noise realizations}
\label{sec:appendix}

We have performed all our simulations in three different noise realizations. To avoid plots getting
too busy, 
in Sec.~\ref{sec:results} we only showed results for one of these; here we also give them for the
other two noise realizations. 

\begin{figure*}[htpb]
\includegraphics[width=0.45\textwidth]{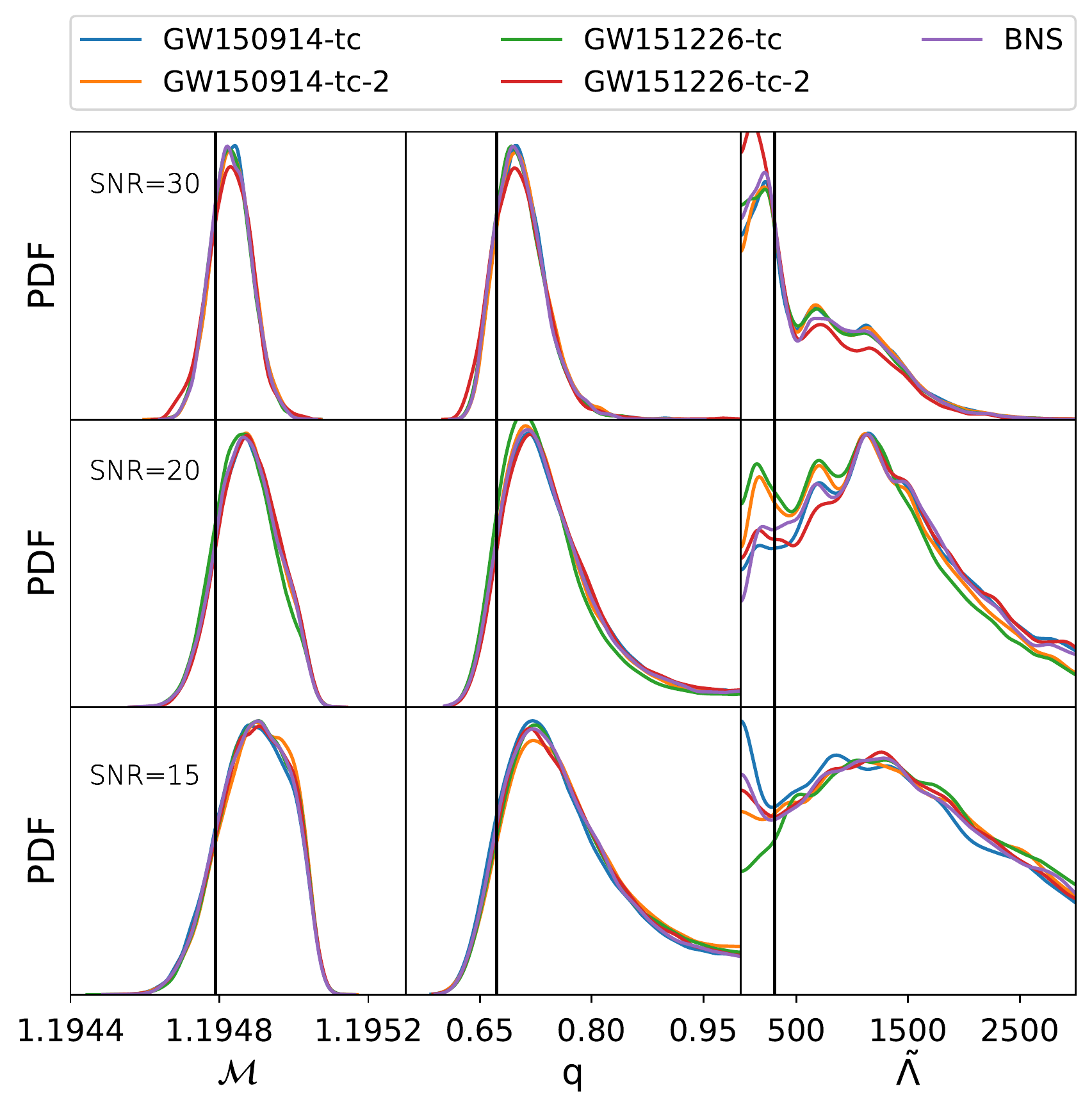}
\includegraphics[width=0.51\textwidth]{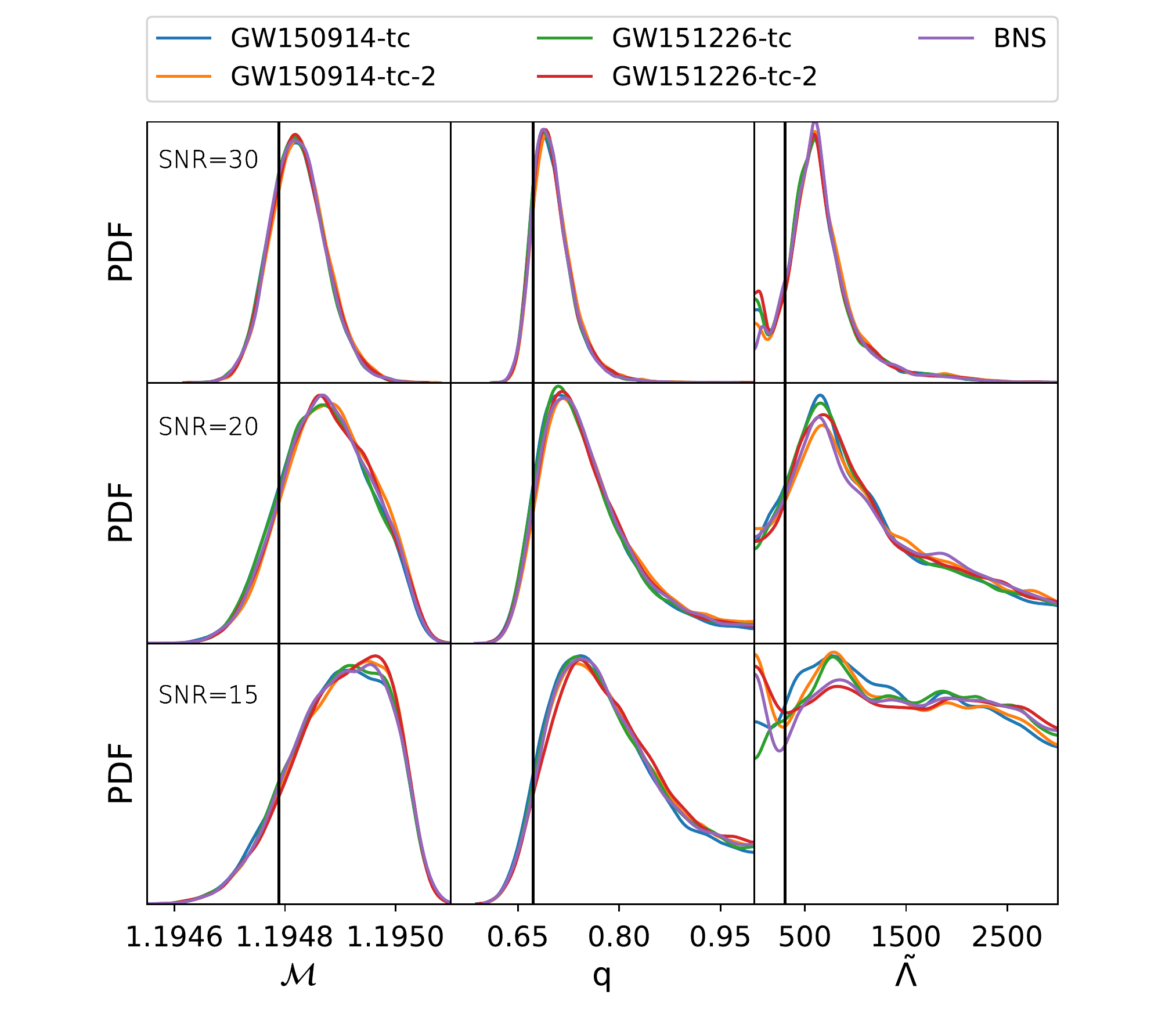}
 \caption{Posterior PDFs for BNS parameters when a BNS and BBH signal are being overlapped; same as 
                     Fig.~\ref{fig:intrinsic_overlap_1} when injections are done in two other noise realizations 
                     (left and right panels).}
 \label{fig:noise23_rec_bns_bbh_plus_bns_inj}
 \end{figure*}

\begin{figure*}[htpb]
\includegraphics[keepaspectratio, width=0.48\textwidth]{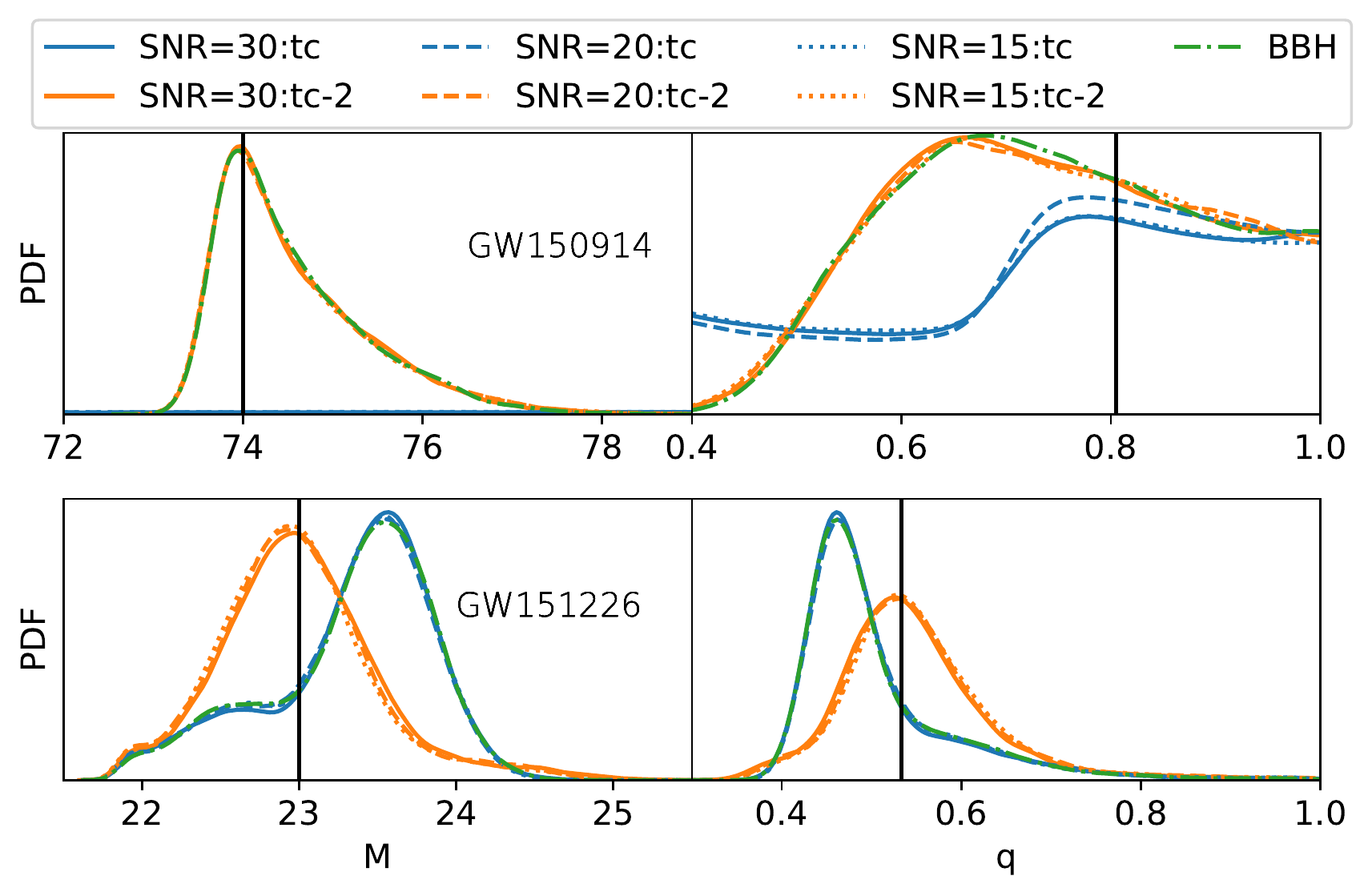}
\includegraphics[keepaspectratio, width=0.48\textwidth]{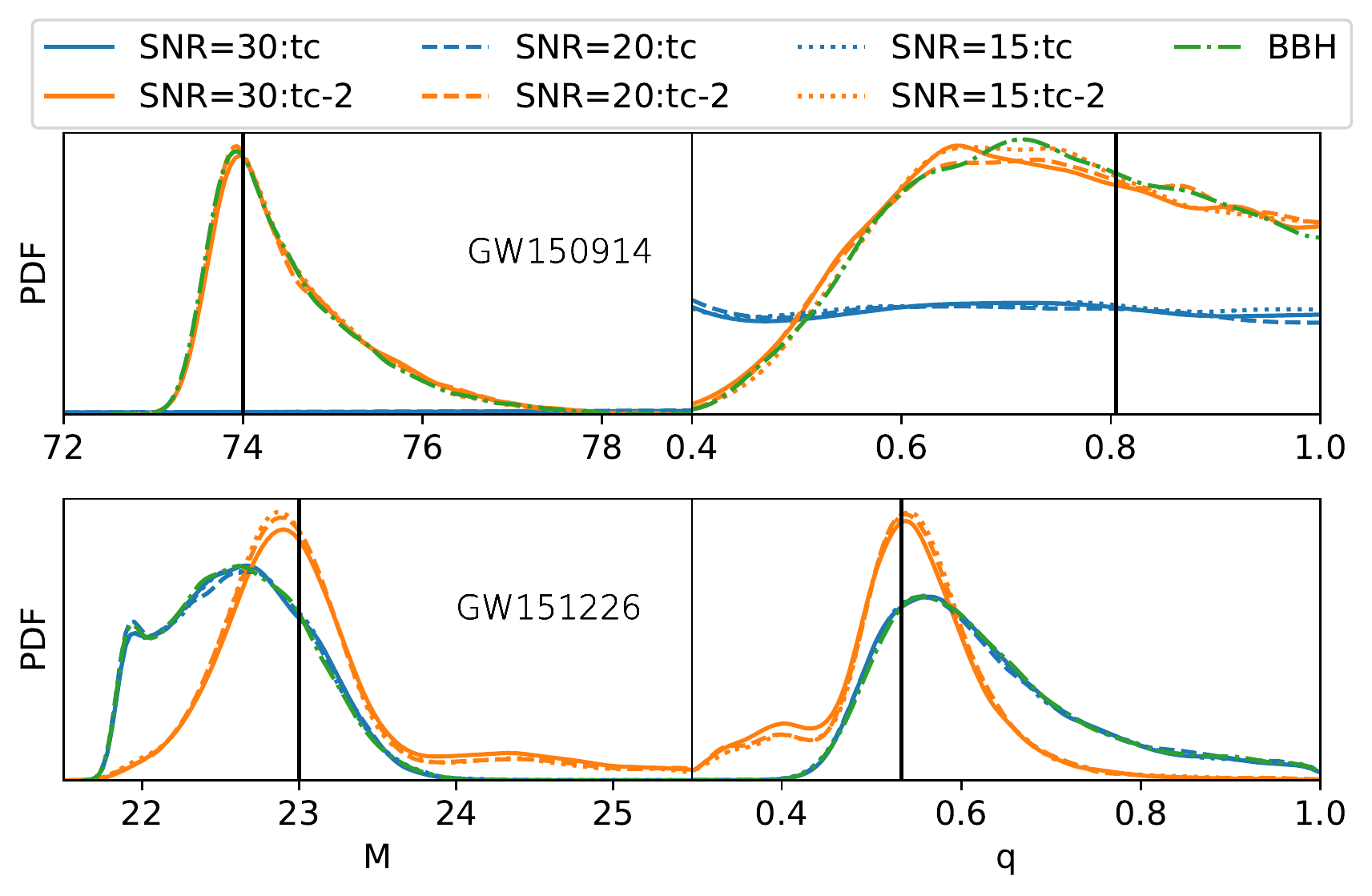}
\caption{Posterior PDFs for BBH parameters when a BNS and BBH signal are being overlapped; same as 
                    Fig.~\ref{fig:intrinsic_overlap_12} when injections are done in two other noise realizations.}
 \label{fig:noise23_rec_bbh_bbh_plus_bns_inj}
 \end{figure*}

In the case of a BNS overlapping with a BBH, the measurements on the BNS  
are shown in Fig.~\ref{fig:noise23_rec_bns_bbh_plus_bns_inj} and those on the BBH in 
Fig.~\ref{fig:noise23_rec_bbh_bbh_plus_bns_inj}. The corresponding KS values 
are given in Tables \ref{tab:KS_BBH_BNS_noise23_left} and \ref{tab:noise23_BBHrec_KS_left}, respectively. 
For measurements on the mass parameters of the 
BNS, we find that the results are consistent between noise realizations. For the
tidal parameter $\tilde{\Lambda}$, the PDFs differ somewhat more; compare the right columns in the 
two panels of Fig.~\ref{fig:noise23_rec_bns_bbh_plus_bns_inj}. This is likely because most of the 
information on tides enters the signal at higher frequencies, where the variance of the noise is 
larger; hence the measurement of $\tilde{\Lambda}$ will be more affected by the noise realization 
than the mass measurements, especially when SNRs are not high. 
Indeed, though not shown here explicitly, KS statistics for $\tilde{\Lambda}$ between different 
noise realizations, but for the same overlap situation,
tend to be significantly larger than within the same noise realization but for different overlaps. 
For parameter estimation on the BBH, there are differences in the PDFs for the masses when the BBH merger time 
coincides with that of the BNS, but not so much if it occurs 2 seconds earlier.

\begin{figure*}[htpb]
\includegraphics[keepaspectratio, width=0.48\textwidth]{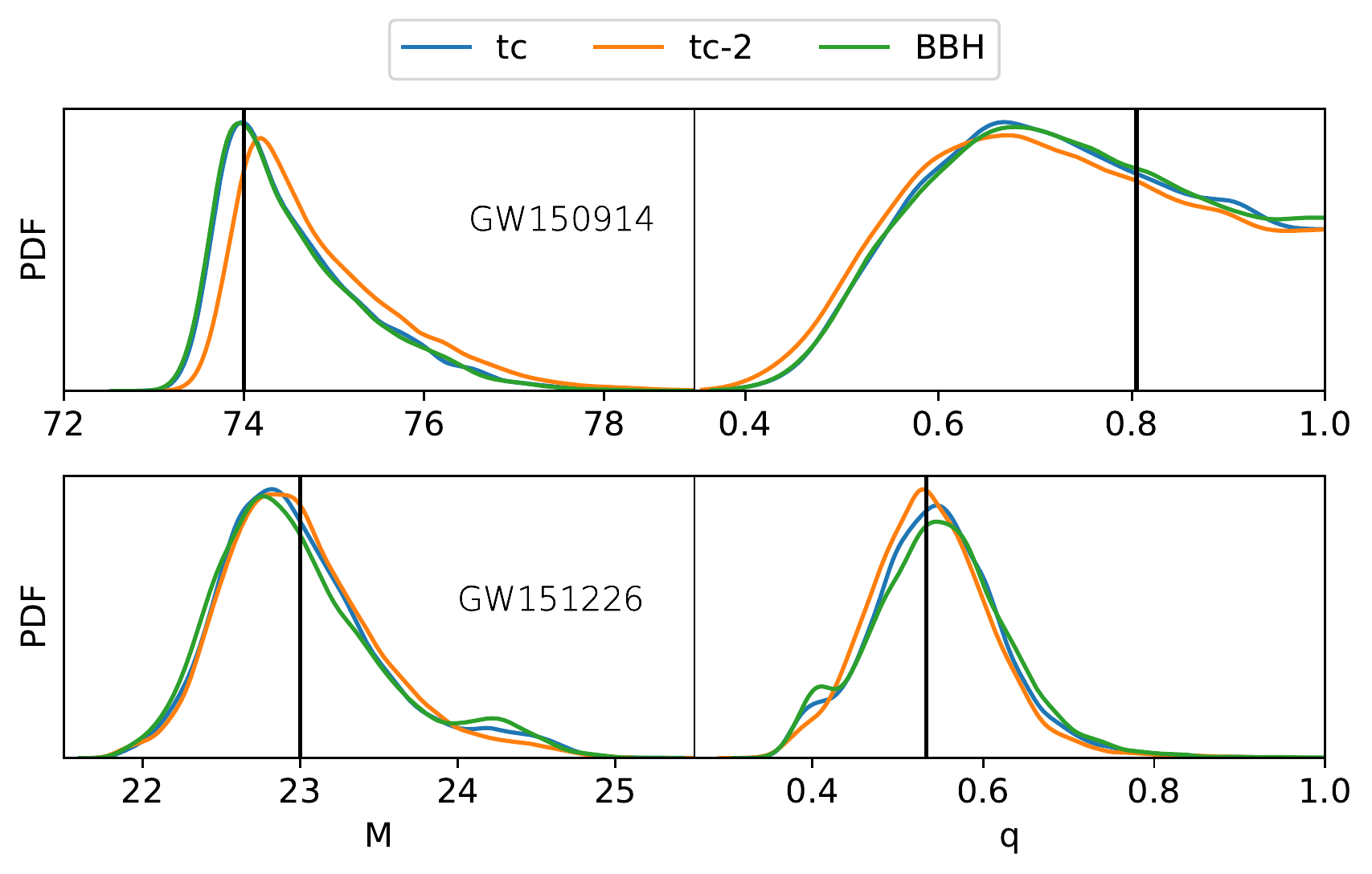}
\includegraphics[keepaspectratio, width=0.48\textwidth]{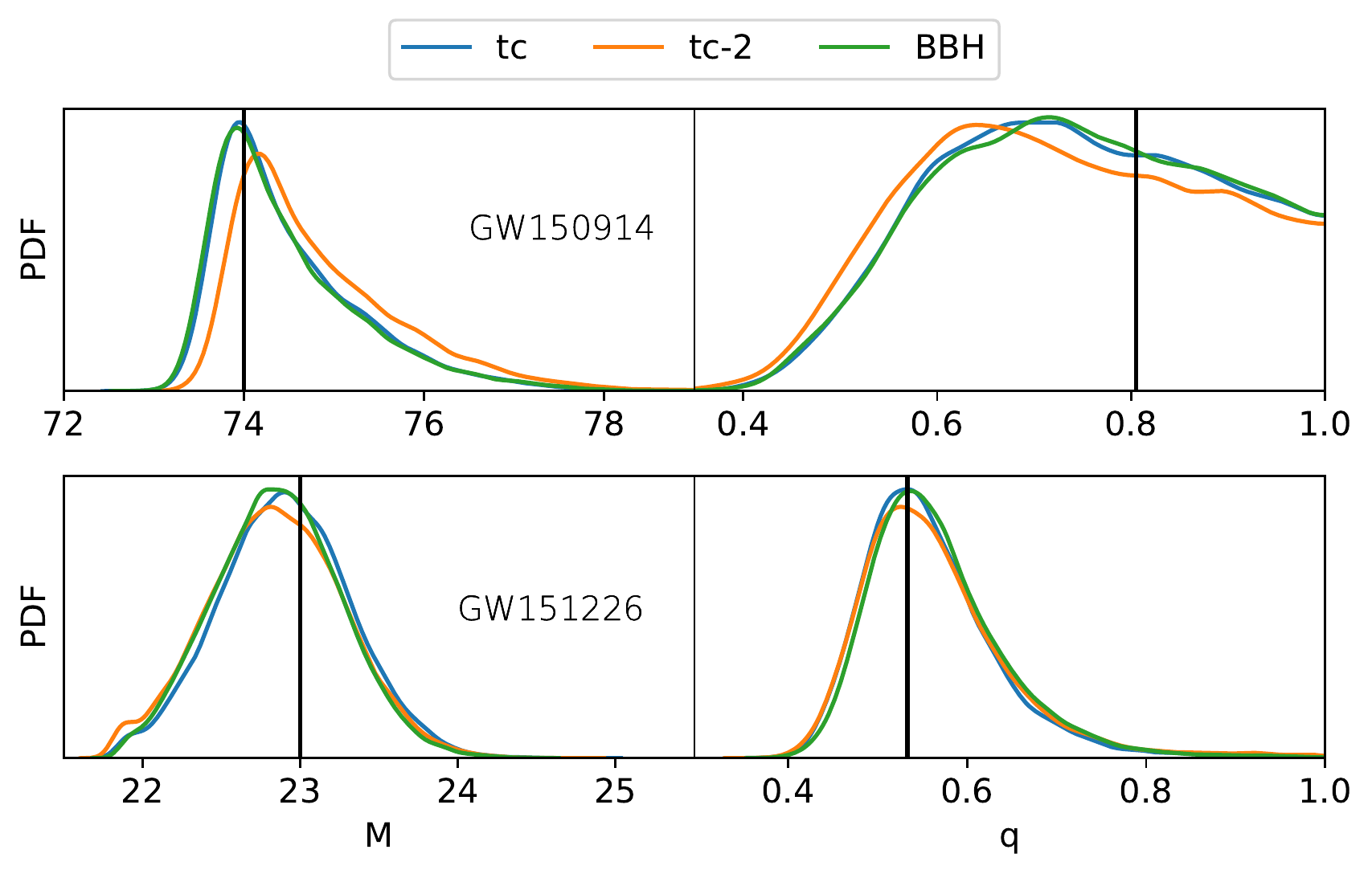}
\caption{Posterior PDFs for BBH parameters when a BNS and BBH signal are being overlapped; same as 
                    Fig.~\ref{fig:2BBHs_noise} when injections are done in two other noise realizations.}
 \label{fig:noise23_rec_bbh_bbh_plus_bbh_inj}
 \end{figure*}

In the case of two overlapping BBH signals, parameter estimation results are  
shown in Fig.~\ref{fig:noise23_rec_bbh_bbh_plus_bbh_inj}, and KS statistics in Table \ref{tab:2BBHs_noise_KS_An1}. The results are quite
robust under a change of noise realization. 

\begin{figure*}[htpb]
\includegraphics[keepaspectratio, width=0.48\textwidth]{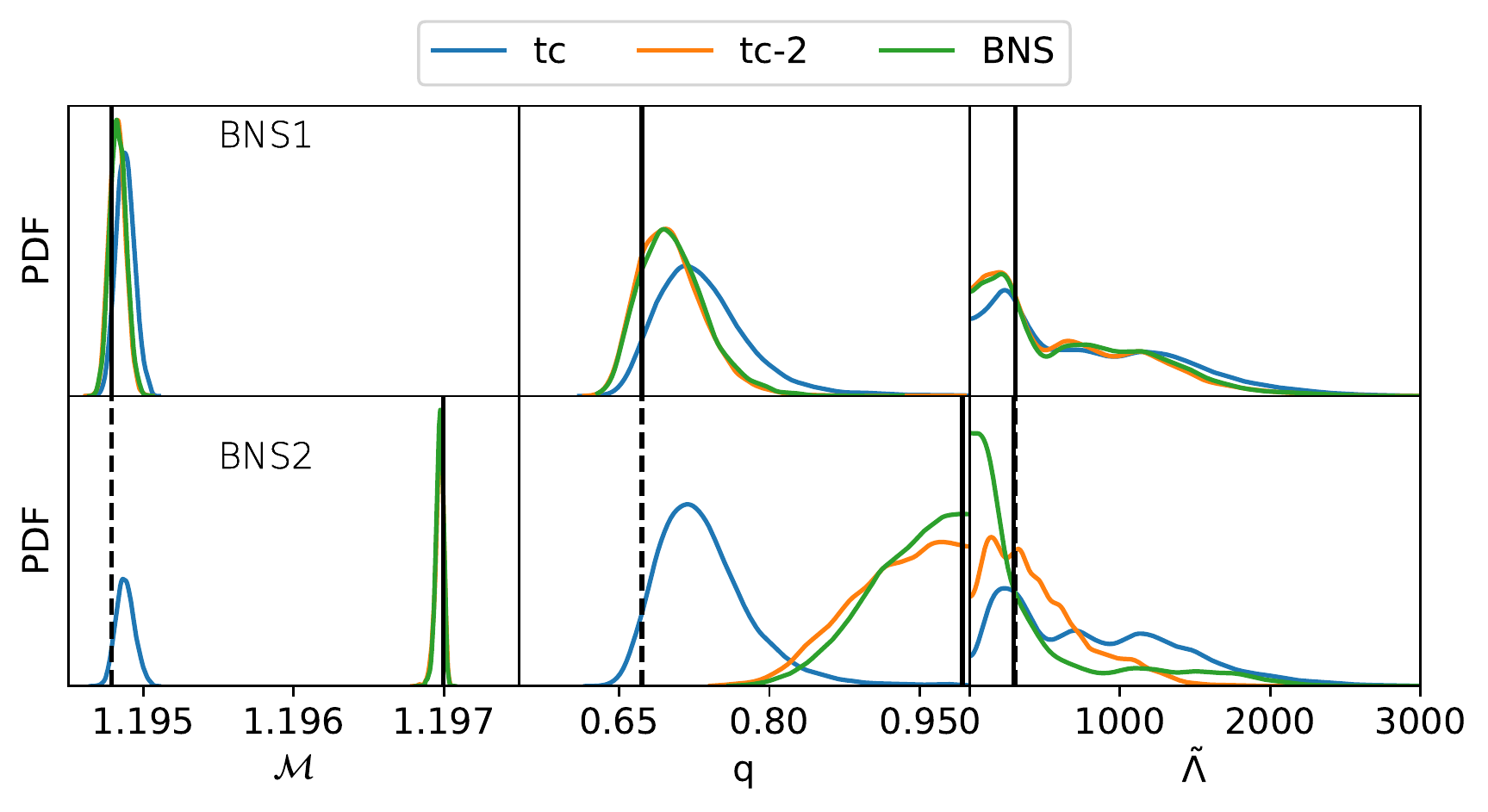}
\includegraphics[keepaspectratio, width=0.48\textwidth]{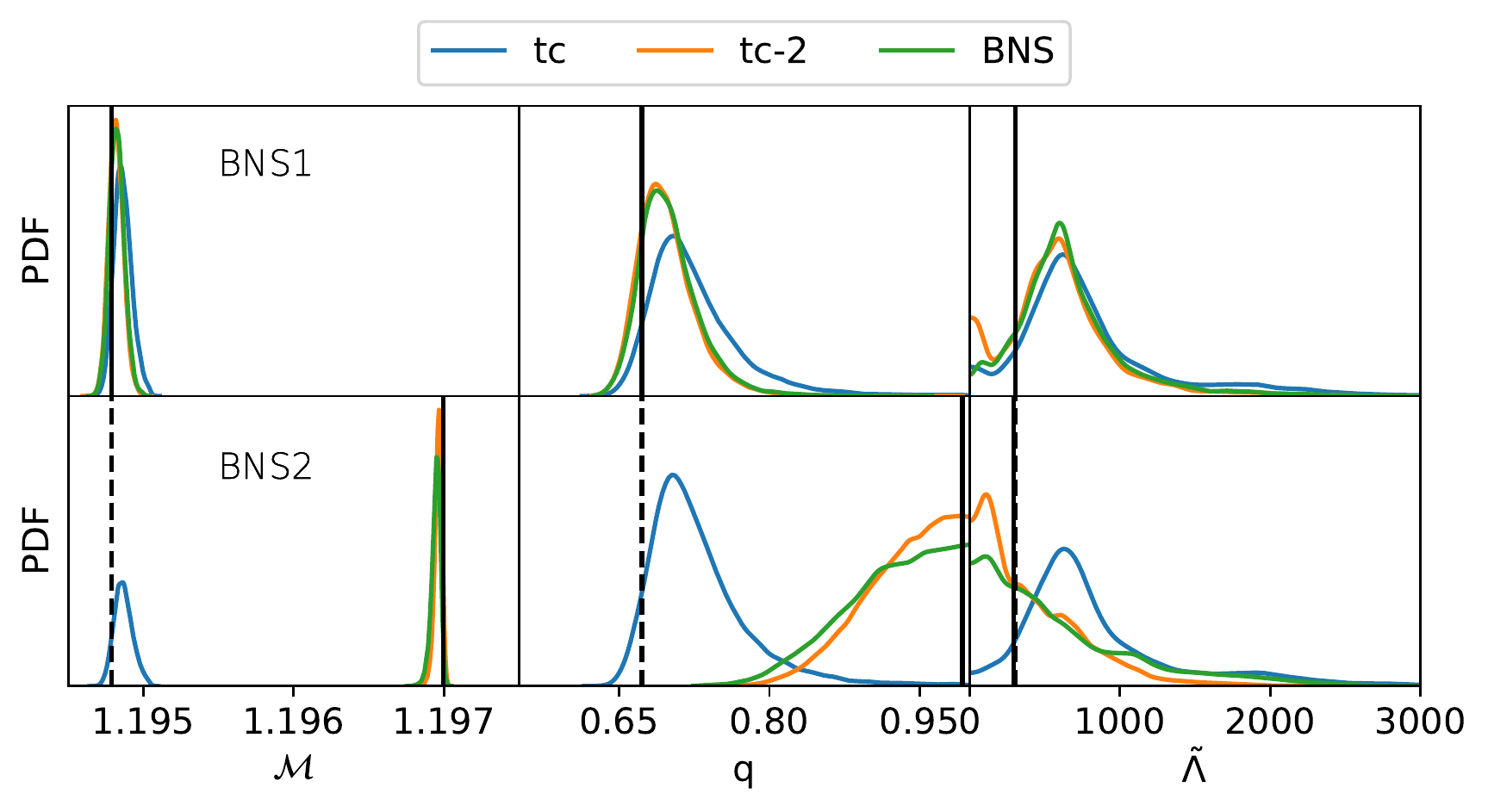}
\caption{Posterior PDFs when two BNS signals are being overlapped; same as 
                    Fig.~\ref{fig:2BNSs_noise} when injections are done in two other noise realizations.}
 \label{fig:noise23_rec_bns_bns_plus_bns_inj}
 \end{figure*}

Finally, the case of two overlapping BNSs with different noise realizations is
shown in Fig.~\ref{fig:noise23_rec_bns_bns_plus_bns_inj}, and KS statistics in Table \ref{tab:2BNS_noise_KS_An1}. 
As in the case of a BNS overlapping
with a BBH, the PDFs for the masses are not much affected by differences in 
noise, but the ones for $\tilde{\Lambda}$ are more susceptible.

 \begin{table*}[htpb]
    \begin{tabular}{| *{10}{c|} }
    \hline
BBH overlapped    & \multicolumn{3}{c|}{BNS (SNR = 30)}
                    & \multicolumn{3}{c|}{BNS (SNR = 20)}
                        & \multicolumn{3}{c|}{BNS (SNR = 15)}  \\
    \hline
   \emph{Noise realization 2} &   $\mathcal{M}$  &   $q$  &   $\tilde{\Lambda}$  &   $\mathcal{M}$  &   $q$  &   $\tilde{\Lambda}$  &   $\mathcal{M}$  &  $q$ & $\tilde{\Lambda}$  \\
   \hline
   \texttt{GW150914-tc} & 0.0267 & 0.0248 & 0.0224 & 0.0106 & 0.0141 & 0.0169 & 0.0146 & 0.0211 & 0.0290 \\
   \hline
   \texttt{GW150914-tc-2} & 0.0287 & 0.0282 & 0.0338 & 0.00601 & 0.0108 & 0.0486 & 0.0263 & 0.0308 & 0.0137 \\
   \hline
   \texttt{GW151226-tc} & 0.0125 & 0.0141 & 0.0421 & 0.0376 & 0.0471 & 0.0723 & 0.0155 & 0.0152 & 0.0333 \\
   \hline
   \texttt{GW151226-tc-2} & 0.0337 & 0.0346 & 0.0815 & 0.0244 & 0.0258 & 0.0179 & 0.0113 & 0.0108 & 0.00923 \\
    \hline
    \hline
   \emph{Noise realization 3} &    $\mathcal{M}$ &  $q$   &   $\tilde{\Lambda}$  &  $\mathcal{M}$ &   $q$   &   $\tilde{\Lambda}$  &  $\mathcal{M}$  &  $q$   & $\tilde{\Lambda}$  \\
   \hline
   \texttt{GW150914-tc} & 0.0140 & 0.0143 & 0.0251 & 0.0236 & 0.0298 & 0.0481 & 0.0114 & 0.0255 & 0.0378  \\
   \hline
   \texttt{GW150914-tc-2} & 0.0296 & 0.0396 & 0.0255 & 0.0272 & 0.0218 & 0.0125 & 0.0186 & 0.0125 & 0.0299  \\
   \hline
   \texttt{GW151226-tc} & 0.0135 & 0.0161 & 0.0347 & 0.0215 & 0.0312 & 0.0412 & 0.00750 & 0.00868 & 0.0239  \\
   \hline
  \texttt{GW151226-tc-2} & 0.0142 & 0.0140 & 0.0334 & 0.0109 & 0.00833 & 0.0310 & 0.0223 & 0.0292 & 0.0169 \\
    \hline
    \end{tabular}
    \caption{Values of the KS statistic comparing PDFs for BNS parameters (columns) in the BNS+BBH overlap scenarios (rows) with the 
    corresponding PDFs when there is no overlapping BBH signal, when injections are done in two other noise realizations.
    The numbers shown correspond to the PDFs in Fig.~\ref{fig:noise23_rec_bns_bbh_plus_bns_inj}, 
    noise realisation 2 corresponding to the left panel and noise realisation 3 to the right panel.}
    \label{tab:KS_BBH_BNS_noise23_left}
\end{table*}

 \begin{table*}[htpb]
    \begin{tabular}{| *{9}{c|} }
    \hline
BNS overlapped   & \multicolumn{2}{c|}{\texttt{GW150914-tc}}
                    & \multicolumn{2}{c|}{\texttt{GW150914-tc-2}}
                        & \multicolumn{2}{c|}{\texttt{GW151226-tc}}
                           & \multicolumn{2}{c|}{\texttt{GW151226-tc-2}} \\
    \hline
   \emph{Noise realization 2}  &   $M$  &   $q$  &   $M$   &   $q$   &   $M$ &   $q$  &   $M$  &   $q$  \\
   \hline
   BNS (SNR = 15) & -- & -- & 0.0134 & 0.011 & 0.00832 & 0.00890 & 0.411 & 0.398   \\
   \hline
   BNS (SNR = 20) & -- & -- & 0.0104 & 0.0109 & 0.0169 & 0.0172 & 0.390 & 0.377 \\
   \hline
   BNS (SNR = 30) & -- & -- & 0.0100 & 0.0113 & 0.0140 & 0.0146 & 0.367 & 0.357  \\
   \hline
   \hline
    \emph{Noise realization 3} &   $M$  &  $q$  &   $M$   &  $q$   &   $M$  &  $q$   &   $M$  &  $q$  \\
   \hline
   BNS (SNR = 15) & -- & -- & 0.0168 & 0.0100 & 0.0140 & 0.0142 & 0.318 & 0.131  \\
   \hline
   BNS (SNR = 20) & -- & -- & 0.0189 & 0.0131 & 0.0132 & 0.0137 & 0.322 & 0.315 \\
   \hline
   BNS (SNR = 30) & -- & -- & 0.0287 & 0.295 & 0.0136 & 0.0130 & 0.334 & 0.327 \\
   \hline
    \end{tabular}
    \caption{Values of the KS statistic comparing PDFs for BBH parameters (columns) in the BNS+BBH overlap scenarios (rows) with the 
    corresponding PDFs when there is no overlapping BNS signal, when injections are done in two other noise realizations.
    As before, when the GW150914-like signal ends at the same time as a BNS, it is not found by the sampling algorithm, but
    other scenarios are less problematic. The numbers shown correspond to the PDFs in Fig.~\ref{fig:noise23_rec_bbh_bbh_plus_bns_inj}, 
    noise realisation 2 corresponding to the left panel and noise realisation 3 to the right panel.}
    \label{tab:noise23_BBHrec_KS_left}
\end{table*}

\begin{table*}[htpb]
    \begin{tabular}{| *{8}{c|} }
    \hline
    \multicolumn{2}{|c|}{\texttt{GW150914-tc}}
                    & \multicolumn{2}{c|}{\texttt{GW150914-tc-2}}
                        & \multicolumn{2}{c|}{\texttt{GW151226-tc}}
                           & \multicolumn{2}{c|}{\texttt{GW151226-tc-2}} \\
    \hline
    $M$  &   $q$  &   $M$  &   $q$ &   $M$  &  $q$  &   $M$  &   $q$  \\
   \hline
    0.0195 & 0.00854 & 0.163 & 0.0395 & 0.0299 & 0.0309 & 0.0417 & 0.0746  \\
   \hline
   \hline
   $M$  &   $q$  &   $M$  &   $q$  &   $M$  &   $q$  &   $M$  &   $q$  \\
   \hline
    0.0291 & 0.0110 & 0.188 & 0.0625 & 0.0477 & 0.0497 & 0.0225 & 0.0440 \\
   \hline
    \end{tabular}
    \caption{Values of the KS statistic comparing PDFs for BBH parameters in the BBH+BBH overlap scenarios with the 
    corresponding PDFs for the BBH-only case, when injections are done in two other noise realizations. 
    The numbers shown correspond to the PDFs in Fig.~\ref{fig:noise23_rec_bbh_bbh_plus_bbh_inj}, 
    the upper row corresponding to the left panel and the lower row to the right panel.}
    \label{tab:2BBHs_noise_KS_An1}
\end{table*}

\begin{table*}[htpb]
    \begin{tabular}{| *{12}{c|} }
    \hline
\multicolumn{3}{|c|}{BNS1 (\texttt{tc})}
                      & \multicolumn{3}{c|}{BNS1 (\texttt{tc-2})}
                        & \multicolumn{3}{c|}{BNS2 (\texttt{tc})}  
                            & \multicolumn{3}{c|}{BNS2 (\texttt{tc-2})}\\
    \hline
   $\mathcal{M}$  &   $q$  &   $\tilde{\Lambda}$  &   $\mathcal{M}$  &   $q$  &   $\tilde{\Lambda}$  &   $\mathcal{M}$  &  $q$ & $\tilde{\Lambda}$ &   $\mathcal{M}$  &  $q$ & $\tilde{\Lambda}$  \\
   \hline
   0.316 & 0.282 & 0.0743 & 0.0385 & 0.0339 & 0.0325 & 1.0 & 0.936 & 0.382 & 0.0271 & 0.0858 & 0.248 \\
   \hline
   \hline 
   $\mathcal{M}$  &  $q$  &   $\tilde{\Lambda}$  &  $\mathcal{M}$  &  $q$  &   $\tilde{\Lambda}$  &  $\mathcal{M}$  &  $q$ & $\tilde{\Lambda}$ &  $\mathcal{M}$  &  $q$ & $\tilde{\Lambda}$  \\
   \hline
   0.278 & 0.257 & 0.123 & 0.0475 & 0.0381 & 0.0630 & 1.0 & 0.902 & 0.341 & 0.226 & 0.101 & 0.128 \\
    \hline
    \end{tabular}
    \caption{Values of the KS statistic comparing PDFs for BNS parameters in the BNS+BNS overlap scenarios with the 
    corresponding PDFs for the BNS-only case, when injections are done in two other noise realizations.
    The numbers shown correspond to the PDFs in Fig.~\ref{fig:noise23_rec_bns_bns_plus_bns_inj}, the 
    upper row corresponding to the left panel and lower row to the right panel.}
    \label{tab:2BNS_noise_KS_An1}
\end{table*}

\end{document}